\documentclass[pre,aps,superscriptaddress,twocolumn,amsmath,amssymb]{revtex4-1} 

\usepackage{hyperref}
\usepackage{graphics}
\usepackage{epsfig}
\usepackage{epstopdf}
\usepackage{amsmath}
\usepackage{setspace}
\usepackage{graphicx}% Include figure files
\usepackage{dcolumn}% Align table columns on decimal point
\usepackage{bm}% bold math
\usepackage{color}
\usepackage{latexsym}
\usepackage{textcomp}

\begin{document}

\title{Arrested States in Persistent Active Matter: Gelation without Attraction}

\author{Carl Merrigan}

\affiliation{Martin Fisher School of Physics, Brandeis University, Waltham, MA 02454, USA}

\author{Kabir Ramola}

\affiliation{Centre for Interdisciplinary Sciences, Tata Institute of Fundamental Research, Hyderabad 500107, India}

\author{Rakesh Chatterjee}

\affiliation{School of Mechanical Engineering and Sackler Center for Computational Molecular and Materials Science, Tel Aviv University, Tel Aviv 69978, Israel}

\author{{Nimrod Segall}}

\affiliation{School of Mechanical Engineering and Sackler Center for Computational Molecular and Materials Science, Tel Aviv University, Tel Aviv 69978, Israel}

\author{Yair Shokef}

\affiliation{School of Mechanical Engineering and Sackler Center for Computational Molecular and Materials Science, Tel Aviv University, Tel Aviv 69978, Israel}

\author{Bulbul Chakraborty}

\affiliation{Martin Fisher School of Physics, Brandeis University, Waltham, MA 02454, USA}

\begin{abstract}

We explore phase separation and kinetic arrest in a model active colloidal system consisting  of  self-propelled, hard-core particles with  non-convex shapes.  The passive limit  of the model, namely cross-shaped particles on a square lattice, exhibits a first order transition from a fluid phase to a solid phase with increasing density. Quenches into the two-phase coexistence region exhibit an aging regime. 
The non-convex shape of the particles eases jamming in the passive system and leads to strong inhibition of rotations of the active particles.  Using numerical simulations and analytical modeling,  we quantify the non-equilibrium phase behavior as a function of density and activity.  If we view activity as the analog of attraction strength, the phase diagram exhibits strong similarities to that of attractive colloids, exhibiting both aging, glassy states and gel-like arrested states. The two types of dynamically arrested states, glasses and gels, are distinguished by the appearance of density heterogenities in the latter.  In the infinitely persistent limit, we show that a coarse-grained model based on the asymmetric exclusion process quantitatively predicts the density profiles of the gel states.  The predictions remain qualitatively valid for finite rotation rates. Using these results, we classify the activity-driven phases and identify the boundaries separating them.

\end{abstract}

\maketitle

\section{Introduction}

{
Active matter, constituted of particles that convert ambient energy to directed motion, has emerged as an  important class of non-equilibrium systems with examples ranging from bacterial suspensions to synthetic colloids.   Being driven out of equilibrium at microscopic scales, the collective dynamics of these systems are far richer~\cite{toner2005hydrodynamics,narayan2007long,dey2012spatial,wensink2012meso,theurkauff2012dynamic,palacci2013living,buttinoni2013dynamical} than thermal systems, which are bound by fluctuation-dissipation relations.    

A particular collective behavior that has been widely studied is motility-induced phase separation (MIPS)~\cite{cates2015motility,fily2012athermal,redner2013structure,whitelam2018phase,cugliandolo2017phase,digregorio2018full}.
%
%
% exhibit diverse collective dynamics that would be impossible for thermalized particles, such as ordered flocking patterns ~\cite{toner2005hydrodynamics}, giant number fluctuations ~\cite{narayan2007long,dey2012spatial}, turbulent motion ~\cite{wensink2012meso}, dynamical clustering ~\cite{theurkauff2012dynamic, palacci2013living} and phase separation in the absence of any attractive interparticle forces ~\cite{buttinoni2013dynamical}. 
% 
% In particular, a large amount of research has focused on elucidating the kinetic origins of phase separation of self-propelled particles~\cite{fily2012athermal,redner2013structure,whitelam2018phase}, usually called ``Motility-Induced-Phase-Separation" (MIPS) ~\cite{cates2015motility}. 
% 
MIPS, a kinetic phenomenon,  is striking in its similarity to  equilibrium  phase separation  such as in passive colloids with attractive interactions~\cite{Zaccarelli_2007}. The non-equilibrium phases and  transitions between them while having exact analogs in equilibrium systems~\cite{cates2015motility,fily2012athermal,redner2013structure,whitelam2018phase,cugliandolo2017phase,digregorio2018full} exhibit anomalous fluctuations that can be traced back to their non-equilibrium nature.   The universality of MIPS has led to the proposition that activity mimics attraction~\cite{redner2013structure}.  Under certain conditions, the non-Brownian random walks representing active-particle dynamics can be mapped onto systems with detailed balance~\cite{PhysRevLett.100.218103}.  

In addition to phase separation, passive colloids exhibit dynamically arrested phases in the form of glasses and gels.  These two types of disordered, amorphous solids have distinct structural and dynamical signatures.  The glass transition occurs in both repulsive and attractive colloids at packing fractions close to  random close packing and is structurally homogeneous on large length scales.  Gelation in attractive colloids leads to strongly heterogeneous states with fluid-like regions coexisting with an arrested, percolated, dense phase~\cite{Zaccarelli_2007}. Active analogs of the glass transition~\cite{Berthier:2013aa,berthier2014nonequilibrium,ni2013pushing,szamel2015glassy,szamel2015glassy,C6SM01322H,berthier2017active,berthier2019perspective,nandi2017nonequilibrium,Nandi7688} and jamming~\cite{fily2014freezing,PhysRevE.84.040301} have been explored extensively in active Brownian particles (ABPs) interacting via repulsive potentials. A review of the emergent behavior of active particles in crowded environments appears in~\cite{RevModPhys.88.045006}. Recent work on an extreme limit of ABPs with long persistence time of their self-propulsion direction has revealed  fluctuations in the dense limit that are qualitatively different from those at short persistence times~\cite{mandal2019extreme}. In this limit of long, yet finite persistence times, clustering and heterogeneous dynamics analogous to passive gels has  been observed~\cite{levis2014clustering} lending further credence to the idea that activity can act as an effective attractive interaction. Similarly, soft disks with infinitely persistent active motion along quenched random directions also display a transition to an absorbing jammed phase above a critical density    \cite{reichhardt2014absorbing}. Other studies have focussed on the variation of persistence times in such systems using the static fluid structure as well as non-equilibrium velocity correlations \cite{szamel2015glassy,C6SM01322H,berthier2017active,berthier2019perspective}. The activity-induced change in the effective attraction in these systems depends on the microscopic details of the particles as well as their persistence times, and therefore activity may enhance or suppress glassy dynamics.

%%%%%%%%%%%%%%%%%%%%%%%%%%%%%%%%%%%%%%%%%%%
%%%%%%%%%%%%%%%%%%%%%%%%%%%%%%%%%%%%%%%%%%%
\begin{figure*}
  \includegraphics[width=1\textwidth]{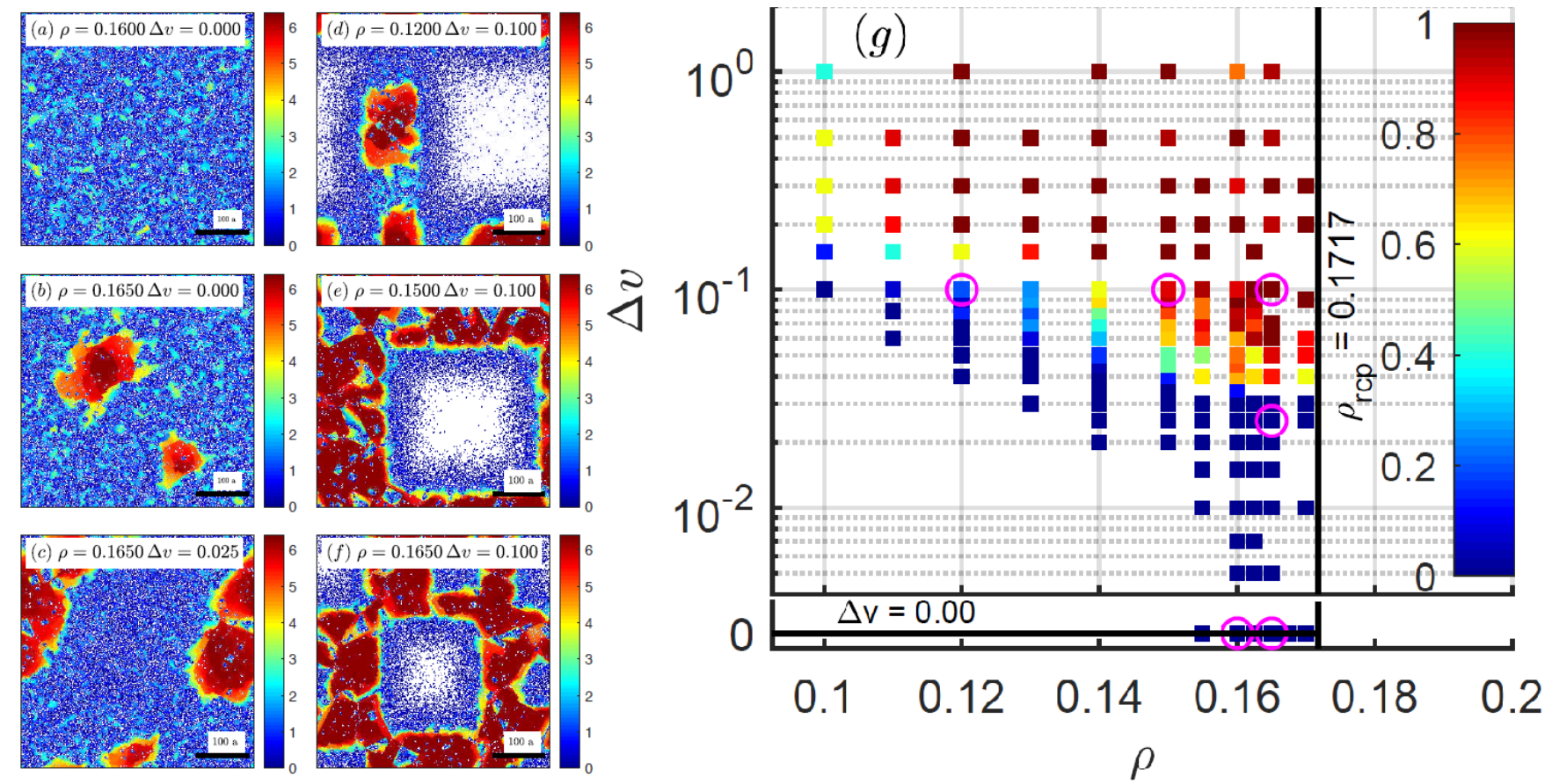} 
\caption{\label{fig1} (a)-(f) Snapshots of the system obtained at the end of  simulations runs,  $ t_{\textrm{max}} = 2 \times 10^6$. The color-bar represents the stationary time for each particle on log scale, $ \log_{10}(\tau_i (t_{max})) $ (see Section \ref{sec:dyn}). Unoccupied lattice sites are colored white.  Particles that have not taken a single step through the duration of the simulation are colored darkest. The phases we find by varying the density $\rho$ and activity $\Delta v$ are (a) steady state passive fluid, (b) passive fluid-solid/aging glass states, (c) finite-activity states resembling the aging regime of the passive system, (d)-(f) active phase with void-solid coexistence (see Section \ref{sec:ASEP}). There is  a progression from majority non-arrested states (d), to majority arrested states (e)-(f) with increasing density. 
%The arrested states exhibit a densely packed, percolating network of immobile crosses which is punctuated by rectangular vacuum pores. In addition, for the arrested states an active liquid forms an interface between the vacuum and the dense solid regions. 
Panel (g) displays a color map of the fraction of arrested states (see Section \ref{sec:MSD}). 
%The dashed black line marks the cross-over to majority arrested states ($p_{\textrm{arrest}}>0.5$), (see text). 
Locations of the snapshots shown in (a)-(f) are marked on the phase diagram with pink circles. $\rho_{\rm{rcp}}$ in the figure denotes the random-close-packing density of hard crosses: the maximal density that can be reached via the RSAD process~\cite{eisenberg2000random}. The morphology exhibited by the snapshots at high densities (e)-(f) are reminiscent of high-density gels in attractive colloids, while (c)-(d) resemble the gel-bubble states at low densities~\cite{Cates_2004}.}
\end{figure*}
%%%%%%%%%%%%%%%%%%%%%%%%%%%%%%%%%%%%%%%%%%%
%%%%%%%%%%%%%%%%%%%%%%%%%%%%%%%%%%%%%%%%%%%

In this paper, we explore the two well-known paradigms of dynamical arrest in passive colloids, gelation and glass formation, in a lattice model of ABPs  with {\it purely repulsive} interactions but with a non-convex shape that can interlock and hinder rotations.
%and show that both these forms of non-equilibrium solidification are realized in ABPs in the limit of infinite persistence times  or  strong rotational locking. 
Using numerical simulations and a coarse-grained model based on a mapping to an asymmetric simple exclusion process (ASEP)~\cite{lebowitz1988microscopic}, we classify the activity-induced phases and construct a non-equilibrium phase diagram.
%We construct a non-equilibrium phase diagram of the {\color{blue} activity-induced arrested states
%using numerical simulations and analytic calculations based on a mapping to an asymmetric simple exclusion process (ASEP)~\cite{lebowitz1988microscopic}.  
A novel feature of this phase diagram is the appearance of a phase with coexisting voids and solids separated by an interface wetted by an active fluid.  The coarse-grained model quantitatively predicts the width of this interface,  which combined with conservation laws leads to predictions of the non-equilibrium phase boundaries separating the arrested states. The main result, summarized in Fig. \ref{fig1}, is that activity triggers arrest into a percolated phase of immobile particles akin to a gel in attractive colloids.  At high densities, the transition is from an aging glass, whereas at low densities it is from an active fluid. To our knowledge, this model provides the first realization of an activity-induced transition from a repulsive glass to a gel.

As described in detail below,  we study an active lattice gas model~\cite{whitelam2018phase,PhysRevLett.120.268003} of hard-core, cross-shaped particles on a square lattice, which is also referred to as the N3 model since each cross prevents the occupation of the first, second and third neighbors of its central square, as shown in Fig. \ref{model_figure}.  In the passive limit, this is the simplest lattice-gas that exhibits a finite-density first-order transition from a fluid phase to a sublattice-ordered phase with tenfold symmetry~\cite{nath2014multiple}.  The sublattice-ordered states can be further grouped into right-handed and left-handed chiral order. In continuum, experiments have demonstrated the emergence of  long-range chiral order in crystals of cross-shaped particles~\cite{Zhao_2014}.  As the density is quenched into the two-phase coexistence region~\cite{eisenberg2000random,eisenberg1998random}, one observes a crossover from a simple fluid to a slowly coarsening or aging  regime in which concentrated immobile clusters with local crystalline order emerge.  There is evidence for the existence of a glass transition~\cite{eisenberg2005first} in this passive system in the form of diverging timescales and the appearance of dynamical heterogeneities~\cite{rotman2009ideal,rotman2010direct}.   

This paper is organized as follows. Section \ref{sec:model} describes our model and the simulation methodology.  In Section \ref{sec:dyn}, we quantify the spatially  heterogeneous dynamics that is visible in the snapshots shown in Fig. \ref{fig1}.   Next, in Section \ref{sec:MSD}, we classify the states into two categories, arrested and non-arrested, based on  measurements of the mean-squared-displacements (MSDs) of the particles.  The MSD measurements also distinguish between aging, glassy states, and gel-like arrested states.  This classification is used to construct the phase diagram shown in Fig. \ref{fig1} (g).  In Section \ref{sec:ASEP}, we present a coarse-grained model of the dynamics that leads to a prediction of the density profile.  We compare these results to  the density profiles obtained in our numerical simulations in Section \ref{sec:density}, and construct a non-equilibrium phase diagram that delineates states based on the density profiles of the arrested states.  This classification connects the dynamical signatures of arrest shown in Section \ref{sec:MSD} to phase separation. Lastly, in Section \ref{sec:finite_rotation}, we consider the effects of finite persistence times. The appendices provide further details of the passive, glassy dynamics, and discuss finite size effects.

%%%%%%%%%%%%%%%%%%%%%%%%%%%%%%%%%%%%%%%%%%%%%%%%%%%%%%%%%%%%%%%%%%%%%%%
\section{Model and Simulations}
\label{sec:model}

%%%%%%%%%%%%%%%%%%%%%%%%%%%%%%%%%%%%%%%%%%%%%%%%%%%%%%%
%%%%%%%%%%%%%%%%%%%%%%%%%%%%%%%%%%%%%%%%%%%%%%%%%%%%%%%
\begin{figure}[t!]
\includegraphics[width=0.48\textwidth]{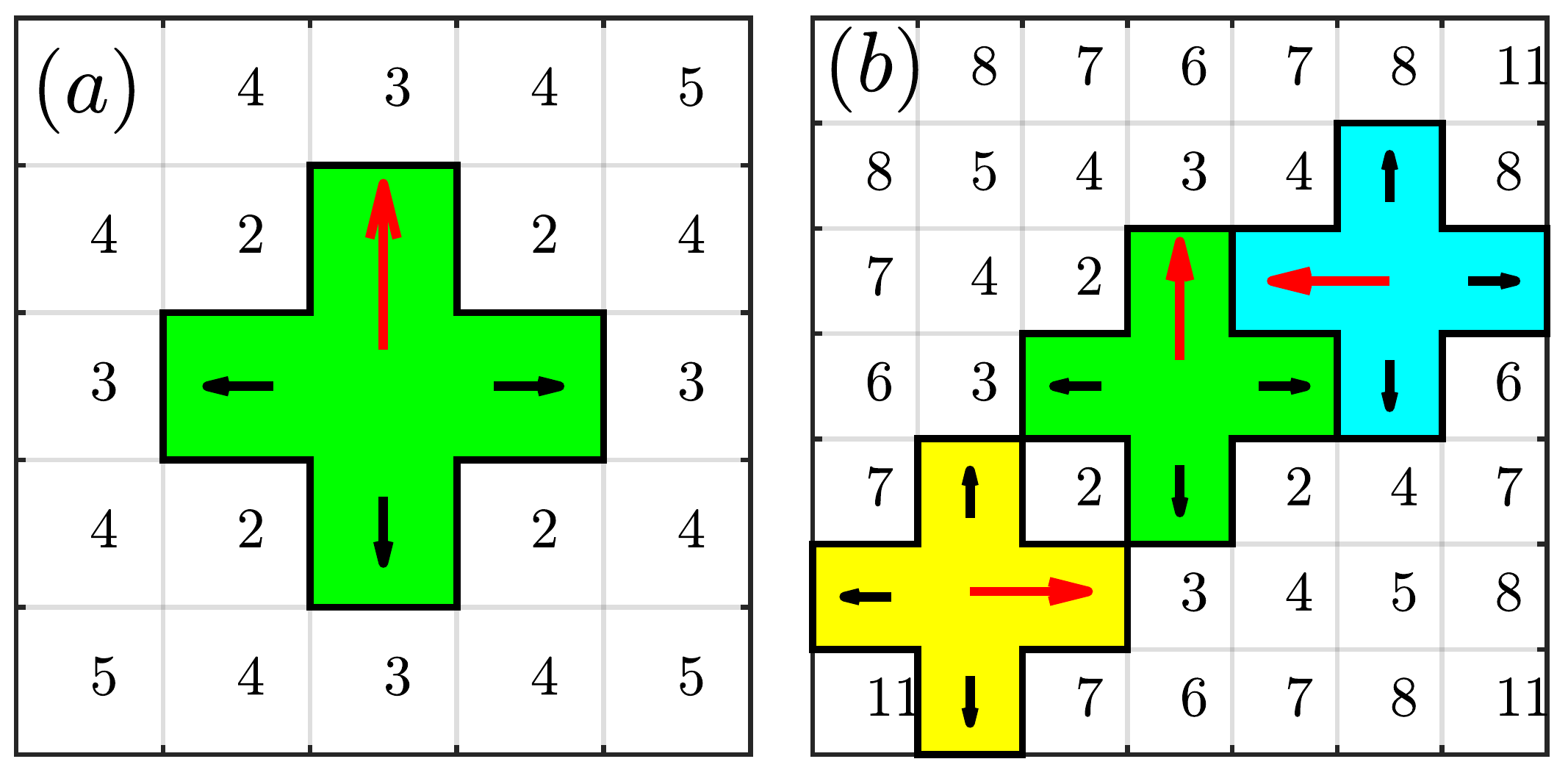}
\caption{(a) Nearest neighbor labels and hopping rates for our active hard-cross model on a square lattice. The particles can perform thermal moves in any of the four directions with a rate $v_0$ (black arrows), and active moves with a rate $v_0 + \Delta v$ along the direction of their orientation (red arrow). (b) The cross at the central site (green) is prevented from rotating by the presence of neighboring crosses occupying the fourth  and fifth nearest-neighbor sites.}
\label{model_figure}
\end{figure}
%%%%%%%%%%%%%%%%%%%%%%%%%%%%%%%%%%%%%%%%%%%%%%%%%%%%%%%
%%%%%%%%%%%%%%%%%%%%%%%%%%%%%%%%%%%%%%%%%%%%%%%%%%%%%%%

We study a model of hard-core particles on the square lattice with exclusion up to third nearest-neighbors~\cite{bellemans1966phase,orban1982phase,eisenberg2005first,fernandes2007monte,nath2014multiple}. Each particle can be represented as a hard, cross-shaped object occupying five lattice sites, see Fig.~\ref{model_figure}. The highest possible density of $\rho = 0.2$ is achieved for a perfectly ordered arrangement of crosses. This equilibrium model is the simplest hard-core exclusion model on a lattice that exhibits a first-order transition from a fluid to solid phase, with coexistence of fluid at density $\rho_{\textrm{fluid}}  \approx 0.16$ with a crystalline solid at $\rho_{\textrm{solid}} \approx 0.19$~\cite{bellemans1966phase,orban1982phase,eisenberg2005first,fernandes2007monte,nath2014multiple}. There are 10 distinct  sublattice orderings  possible for the crystalline packings. The competition between these phases leads to frustration at high densities, and indeed this model displays a glass transition at higher densities ~\cite{rotman2009ideal,rotman2010direct,eisenberg2000random}.  By adding activity we can, therefore, study active analogs of the glass transition and gelation in passive colloids~~\cite{Zaccarelli_2007,Cates_2004,C4SM00199K}.

In the active generalization of the model~\cite{chatterjee2019motion}, each particle is assigned an active direction which can point along any of the four lattice directions $\left( 0, \frac{\pi}{2},\pi,\frac{3 \pi}{2} \right)$. The particles perform active Brownian walks on the lattice with a rate $v_0+ \Delta v$ along the active direction, where  $v_0$ is the ``thermal'' hopping rate along each of the four lattice directions. Each particle can change its active direction by $\pm \frac{\pi}{2}$ with a rotation rate $D_R$. The thermal diffusion coefficient, $ D_{T} = a^2 v_0 $,  where $ a $ is the lattice spacing, is set to unity in our simulations. Since there is no energy scale in this model, the only role of temperature is to set the magnitude of the diffusion coefficient, which simply fixes the unit of time. The expected self-propulsion velocity for a single cross in the dilute limit is $v_p = a \Delta v $~\cite{chatterjee2019motion}, and hence the translational Peclet number is given by $ Pe_{t} = \frac{v_p a}{D_T} = \frac{\Delta v}{v_0} $. Since we fix $v_0=1$, we use $\Delta v$ to represent $Pe_{t}$.

The  active dynamics we prescribe for hard crosses are identical  to those implemented in simulations of MIPS for a simple-exclusion lattice gas model~\cite{whitelam2018phase}: squares on a square lattice, which do not exhibit an equilibrium phase transition or glassy dynamics. Further, we consider rotation of the active direction and require a rigid rotation of the whole cross. Consequently, rotations are disallowed for crosses that have neighboring crosses which occupy either the fourth or fifth nearest-neighbor site, see Fig. \ref{model_figure} (b).  We note that unlike simulations of continuum active dynamics, $D_T$ cannot be set to zero because of a kinetic trap which only exists for random walks on a lattice~\cite{whitelam2018phase}. This is especially true for non-rotating active particles: without thermal moves to free them, non-rotating active particles become immediately trapped upon collision~\cite{biham1992self,soto2014run}. 

We use a continuous time, rejection-free, kinetic Monte Carlo algorithm to implement the active dynamics~\cite{fichthorn1991theoretical,gillespie1976general,BORTZ197510}. All allowed events in the system are assigned a rate, and the relative weight of each rate determines the probability for the event to occur. { Time proceeds by randomly selecting an event, and then advancing the clock by an interval $ -\log (r)/\mathcal{R} $, where $r$ is a uniform random variable, and $\mathcal{R}$ is the sum of all non-zero rates for the allowed events at a give time $t$. The time increments after each event are exponentially distributed with mean $\langle \Delta t \rangle = \frac{1}{\mathcal{R}}$.}  This algorithm is especially efficient for simulations at large densities, where most moves are disallowed by the excluded volume constraint. 

The initial states of the system are prepared using a Random Sequential Adsorption and Diffusion (RSAD) process ~\cite{eisenberg1998random} which can generate disordered packings up to a maximum density of $\rho_{\textrm{rcp}} = 0.1717...$~\cite{eisenberg2005first}, corresponding to the random-close-packing density for hard crosses.
We study a range of global densities between $ \rho = 0.10$  and  $\rho = 0.17$, and a range of activity values $\Delta v$ from $0$ to $1$. Note that even though $\Delta v$ is a dimensionless activity, it may take values larger than unity~\cite{chatterjee2019motion}.The system domain is a two-dimensional square box of linear length $ L  $, periodic boundary conditions, and a fixed total number of particles $N = \rho L^2$. Unless otherwise stated, we present results for  $L = 450$. The longest simulation time is set to $t = 2 \times  10^6$, which is much larger than the $\alpha$-relaxation time at $\rho = 0.16$, $\tau_{\alpha} \approx 100$ (see Appendix \ref{appendix_a}).

In this work we focus primarily on the limit $D_R \to 0$, where the persistence of the active motion becomes infinite. In this limit  the self-propulsion directions of the particles are  ``quenched'' random variables, and we assign these uniformly with {an equal number of the particles} having an active direction along each of the four lattice directions.  Studying active matter in this limit provides us with the opportunity  to probe the strongest departures from equilibrium systems~\cite{mandal2019extreme}, and as we show in Sections \ref{sec:ASEP} and \ref{sec:density}, allows us to construct a coarse-grained model and obtain quantitative estimates of the density profiles observed in the arrested states.
%describe the  of the non-equilibrium phases.
%Thus we consider the scenario in which $ Pe_{r} \rightarrow \infty $, while the remaining control parameter becomes $ Pe_{t} = \frac{v_p a}{D_T} = \frac{\Delta v}{v_0} $.  
Because of the non-convex shape of the particles, we expect that even for finite rotation rates ($D_R > 0$), neighboring crosses should inhibit one another from rotating~\cite{chatterjee2019motion}. Therefore including rotational locking is important for modeling cases in which non-convex self-propelled particles must reorient their body axis in order to change their direction of motion. In the presence of this rotational locking effect,} small but finite $ D_R $ leads to qualitatively similar results as the infinite persistence-time limit, including overall global arrest due to percolating gel-like structures, as shown in Section \ref{sec:finite_rotation}.

\section{Dynamical Heterogeneity and Activity-Induced Aging} 
\label{sec:dyn}

\begin{figure}[t!]
\includegraphics[width=0.5\textwidth]{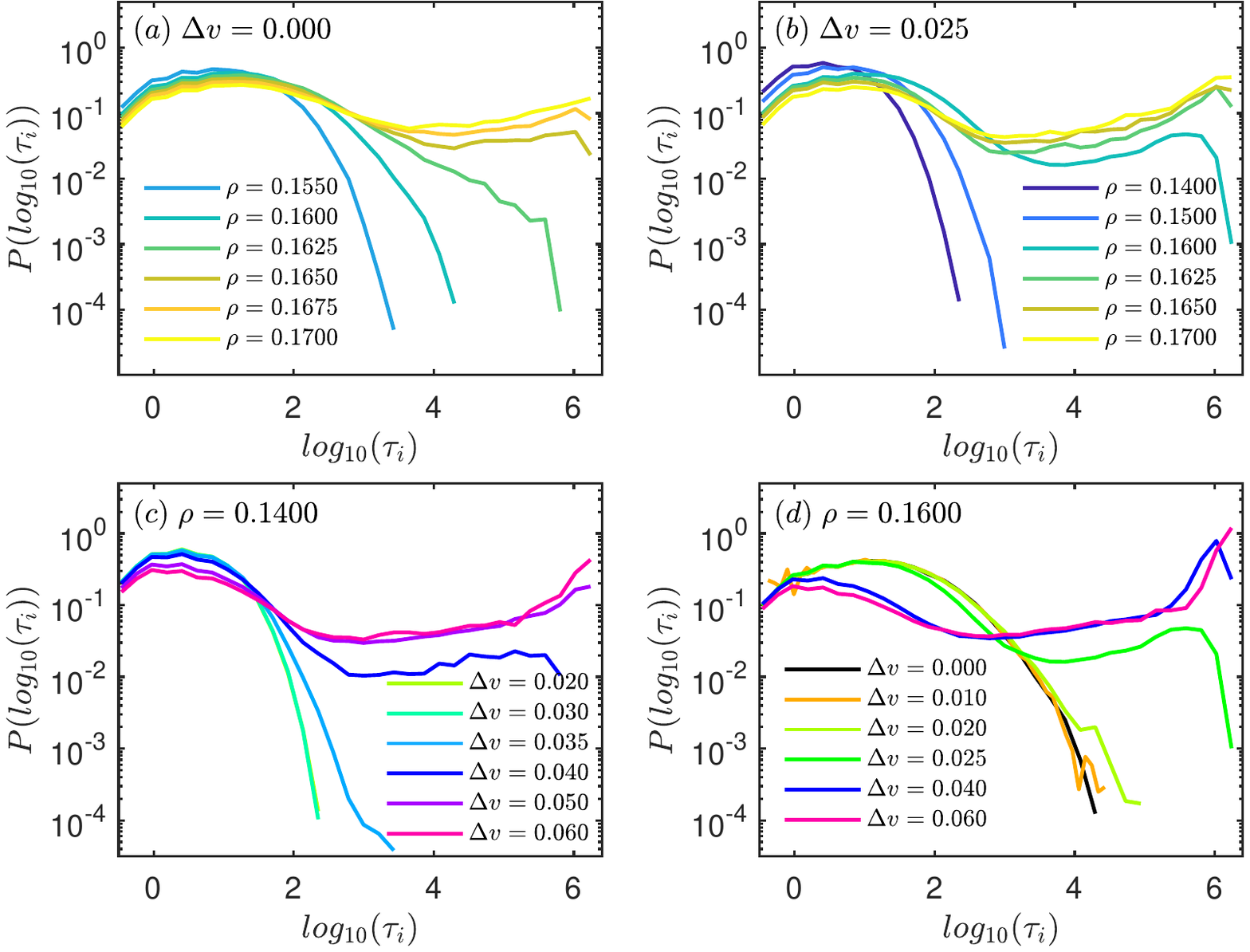}
\caption{Distributions of  stationary times, $ P(\log_{10}(\tau_i (t_{\textrm{max}}))$. In the passive system ($\Delta v = 0$), a signature of the crossover from the steady state passive liquid to an aging glass regime is the appearance of a bimodal distribution, indicating two populations of particles with mobilities that differ by several orders of magnitude. This bimodal distribution appears at lower densities when persistent active motion is introduced. The different figures represent (a) Passive crosses $\Delta v = 0$. (b) The transition to an``active glass" with increasing density at a small activity $\Delta v = 0.025$. The transition at a fixed density (c) $\rho = 0.14$ and (d)  $\rho = 0.16$ with increasing activity.}
\label{fig_stationary_time_distributions} 
\end{figure}

%The snapshots in Fig. \ref{fig1} show  varying degrees of dynamical heterogeneity. 
Our measure of dynamical heterogeneity is based on the definition of ``stationary times" for each cross. The stationary time $\tau_i (t) $ at the observation  time, $t$,  is defined to be the time that cross $i$ has spent at its currently occupied site, $\vec{r}_i$, namely $ \tau_i (t) = t - t_i $, where $ t_i $ is the time at which particle $ i $ arrived at $\vec{r}_i$. Distributions of $\tau_i(t)$ provide a quantitative measure of the spatial heterogeneity of the dynamics, and  is closely related to the distribution of persistence times used to analyze the glass transition in kinetically constrained models~\cite{doi:10.1063/1.2001629,refId0,PhysRevE.92.032133}.   In the aging regime and at large activities, the distributions of $\tau_i$ depend explicitly on time, therefore, 
%We find the stationary times  to be  a good quantitative measure of heterogeneity in arrested states.  
spatial configurations of $\tau_i( t_{\textrm{max}})$ are used to construct the color bar  in Fig.~\ref{fig1}. The distributions of $\tau_i( t_{\textrm{max}})$  are shown in Fig. \ref{fig_stationary_time_distributions}. 

We first discuss the nature of dynamical heterogeneities in the passive system. The adsorption process of the RSAD protocol generates, at time $t=0$,  a uniform configuration with density $ \rho $ in which crystalline order is minimized.  In the passive system, the diffusion of the crosses at $t >0$ leads to equilibrium configurations for $\rho \le 0.1625$.  As shown in Fig. \ref{fig1} (a), these equilibrium states have very little dynamical heterogeneity. The probability distribution of the stationary times, $P(\log_{10}{\tau_i (t_{\textrm{max}}}))$ has a single peak with an increasingly broad tail at large $\tau_i $ as $\rho \rightarrow 0.1625$ (Fig. \ref{fig_stationary_time_distributions} (a)).   For $ \rho > 0.1625 $, the equilibration process is interrupted by an aging/coarsening process in which clusters of increasingly immobile particles grow at an exceedingly slow rate, which prevents the system from reaching a time-translational invariant state. The onset of aging is indicated by the appearance of a bimodal distribution (Fig. \ref{fig_stationary_time_distributions} (a)) and distinct clusters of  particles with  $\tau_i ( t_{\textrm{max}}) \simeq t_{\textrm{max}} $ appearing within a background of particles with $ 10^1 <\tau_i ( t_{\textrm{max}}) < 10^4 $  (Fig. \ref{fig1} (b)). Bimodal distributions of persistence times have been used to identify dynamical heterogeneities in several glass-forming kinetically constrained models~\cite{doi:10.1063/1.2001629,pan2005heterogeneity}. We show in  Appendix \ref{appendix_a} that standard measures such as the self-intermediate scattering function also indicate the onset of aging at $\rho \approx 0.1650$.  
%%%%%%%%%%%%%%%%%%%%%%%%%%%%%%%%%%%%%%%%%%%%%%%%%%%%%%%
%%%%%%%%%%%%%%%%%%%%%%%%%%%%%%%%%%%%%%%%%%%%%%%%%%%%%%%

At non-zero activity, $P(\log_{10}{\tau_i (t_{\textrm{max}}}))$ can develop a bimodal structure at densities lower than $0.165$, as seen in (Figs. \ref{fig_stationary_time_distributions} (b)-(d)). Our results, therefore, indicate that activity leads to a lowering of the onset-density for glassy dynamics, in agreement with the results of  continuum active glass-forming models~\cite{ni2013pushing,mandal2016active}.
%{\color{red} Thus, for increasing density at a finite small activity, glassy dynamics appear at slightly lower density for the active liquid, as has been observed in several continuum active glass-forming models~\cite{ni2013pushing,mandal2016active}.} 
As in the passive, aging  system, these bimodal distributions are associated with states that do not have time-translational invariance.  However, in the active fluids we observe two distinct classes of such states.  At densities $\rho \ge 0.16$, the states (Fig. \ref{fig1} (c)) resemble the passive aging fluid (Fig. \ref{fig1} (b)) with growing clusters of immobile, solid-like regions suspended in a fluid.   Increasing the activity in this density regime leads to states with percolated clusters of immobile particles as seen in  Figs. \ref{fig1} (f).
%Applying an activity $\Delta v = 0.025 $ at densities $\rho \geq 0.16$  leads to the emergence of  immobile clusters  after a characteristic activation time, $\tau_{\textmd{active}}$.  that decreases with increasing density (Fig. \ref{fig_stationary_time_distributions} (b)) and activity (Fig. \ref{fig_stationary_time_distributions} (c) and (d)).   The emergence of  these {\it active}  aging states is indicated by 
%%
%%
%%This behavior indicates that activity leads to aging/coarsening in the density regime where the passive system would usually reach a uniform density steady-state, inducing an aging-state. In Fig. \ref{fig_stationary_time_distributions} (b), we show the appearance of this active aging state, which is characterized by 
%the appearance of a bimodal distribution in $ P(\log_{10}{\tau_i; t_{\textrm{max}}})$, as in the passive case (see Fig. \ref{fig_stationary_time_distributions} (b) and Fig. \ref{fig1} (c)). 
%As $\tau_{\textmd{active}}$ decreases with  stronger activity a system spanning network of immobile crosses develops as shown in Figs. \ref{fig1}(f-g). 
At densities lower than this regime of activity-induced aging,   the appearance of a bimodal distribution in $ P(\log_{10}{\tau_i (t_{\textrm{max}}}))$  with increasing activity  is accompanied by a clear {\it spatial} separation of the particles into ``voids'' and dense regions accommodating the most immobile particles (Fig \ref{fig1} (d)-(e)).   
In this regime, we observe large variations in the final structures from one simulation run to another.     

%%%%%%%%%%%%%%%%%%%%%%%%%%%%%%%%%%%%%%%%%%%%%%%%%%%%%%%
%%%%%%%%%%%%%%%%%%%%%%%%%%%%%%%%%%%%%%%%%%%%%%%%%%%%%%%
\begin{figure}[t!]
\includegraphics[width=0.5\textwidth]{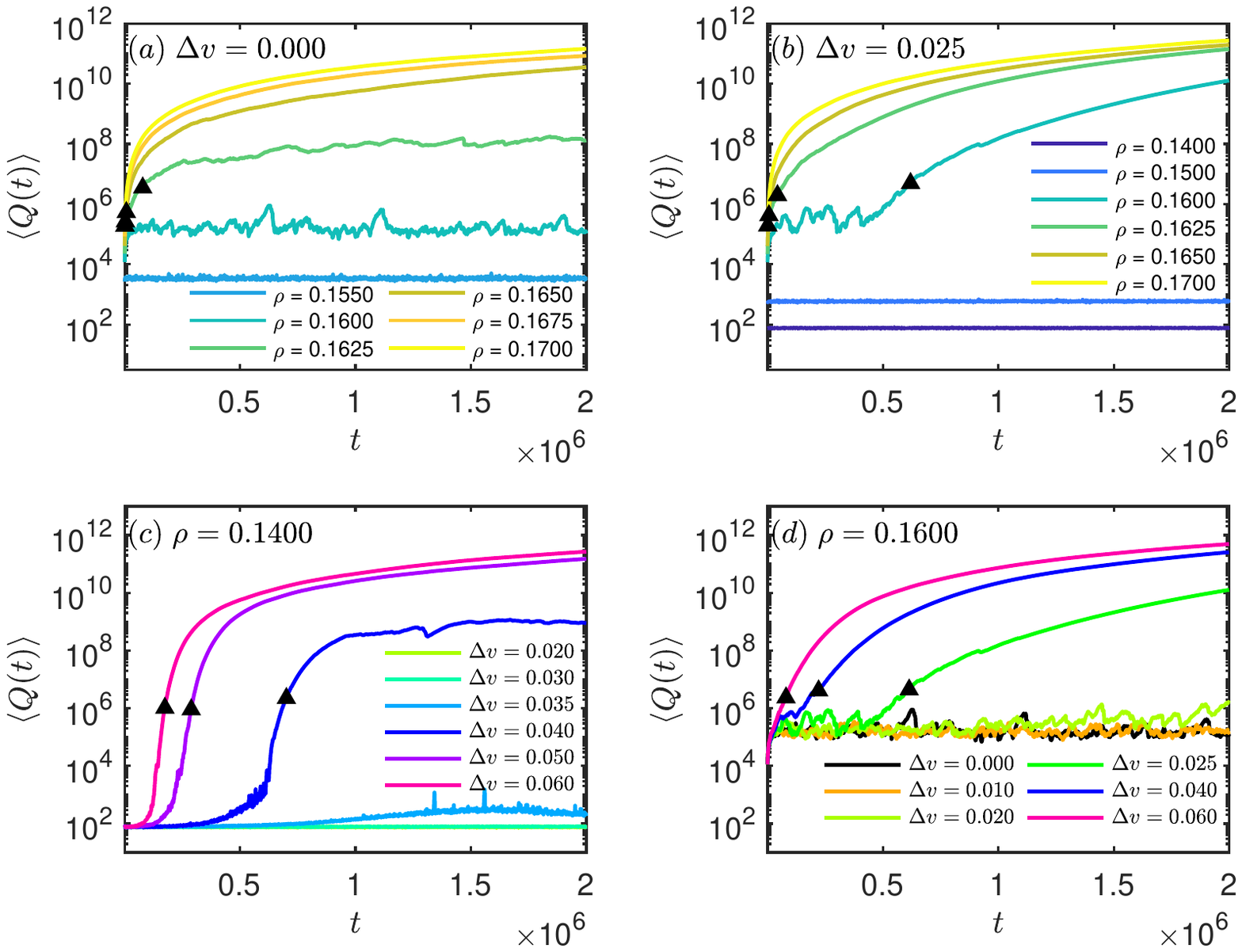}
\caption{Time series of the ensemble averaged stationary time variance, $ \langle Q(t) \rangle $, for the same data sets as in Fig. \ref{fig_stationary_time_distributions}. (a)  Passive system: $ \langle Q(t) \rangle $  increases with time for $ \rho \ge 0.1650 $, indicating the emergence of aging states.  (b) $\Delta v = 0.025$: $ \langle Q(t) \rangle $ increases with time at densities $ \rho \ge 0.1600$.  (c) $ \rho = 0.1400 $:  $ \langle Q(t) \rangle $  increases with time for $\Delta v \ge 0.050$,  (d)$ \rho = 0.1600 $: $ \langle Q(t) \rangle $  increases with time for $\Delta v \ge 0.025$.   The onset time of active dynamics, $\tau_{\textrm{active}}$,  is marked by black arrows in panels (b)-(d).}
\label{fig_variance_stationary_times}
\end{figure}
%%%%%%%%%%%%%%%%%%%%%%%%%%%%%%%%%%%%%%%%%%%%%%%%%%%%%%%
%%%%%%%%%%%%%%%%%%%%%%%%%%%%%%%%%%%%%%%%%%%%%%%%%%%%%%%

The variance of the stationary times, $Q(t) =  \langle (\tau_i (t) - \langle \tau_i (t) \rangle)^2 \rangle $, provides a global measure of the time evolution of spatial heterogeneity in our dynamics. Fig.~\ref{fig_variance_stationary_times} shows the ensemble averaged time series, $\langle Q(t) \rangle$, of $Q(t)$ at different densities and activities. In the passive system, $\langle Q(t) \rangle$ rapidly reaches a small steady state value in the non-aging regime of densities (Fig. \ref{fig_variance_stationary_times} (a)).  Both the equilibration time and the magnitude of the steady-state dynamical heterogeneity grow with density, until for $ \rho > 0.1625 $, the equilibration time surpasses the maximum simulation time.  In this aging regime,  $ \langle Q(t) \rangle  $ grows indefinitely.   
Activity drives strong growth of $\langle Q(t) \rangle$ at lower densities.  However,  as shown in Figs. \ref{fig_variance_stationary_times} (b)-(d), there is a delay time, $ \tau_{\textrm{active}}$, before which the system dynamics match the passive system. To quantify this delay time, we define $\tau_{\textrm{active}}$ to be the time at which the derivative of the stationary-time variance becomes greater than a small threshold, $ d\langle Q(t) \rangle /dt > \epsilon $. For any activity the deriviative is positive for $ t>0 $, but it may remain small for some time: we find $\epsilon = 10^2$ gives a robust signature for the approximate time when the growth rate of $\langle Q(t) \rangle $ first becomes significantly different from zero, see Fig.~\ref{fig_variance_stationary_times}. We have checked that the choice of $\epsilon$  does not significantly change our results other than
providing a scale. This delay time increases with decreasing activity or density.  At very weak activities, therefore,  we observe the passive behavior of $ \langle Q(t) \rangle$ since $\tau_{\textmd{active}}$ increases beyond our maximum simulation time. For instance, at $\rho = 0.1600 $, Fig \ref{fig_variance_stationary_times} (c) shows that the growing dynamical heterogeneity only appears for activities $ \Delta v \geq 0.025$. 

%In the next section, we outline a mean-field theory that argues that  $\tau_{\textmd{active}} \simeq t^*$, the  time required for these non-rotating  active particles to transition from short-time diffusive motion to their eventual ballistic motion at long enough times~\cite{FODOR2018106}.   This timescale, of course, diverges as the activity vanishes, and decreases with increasing density as $1/D(\rho)$, where $D(\rho)$ is the diffusion constant of passive particles at density $\rho$.  At very weak activities, therefore,  we observe the passive behavior of $ \langle Q(t) \rangle$ since $t^*$ ($\tau_{\textmd{active}}$) increases beyond our maximum simulation time. For instance, at $ \rho = 0.1600 $, Fig \ref{fig_variance_stationary_times} (c) shows that the growing dynamical heterogeneity only appears for activities $ \Delta v \geq 0.025 $. 
%{\color{blue} Can we show a scatter plot of $t^*$ vs $\tau_{\textmd{active}}$ in the SI?}

%Therefore, it is possible to make the activity small enough that the basic character of the passive steady state remains unperturbed, even by infinitely persistent active particles. 
%{\color{red} I have rewritten the above because I believe what you are saying is that the time to crossover from diffusive to ballistic exceeds the simulation time at low enough activities.   It would be good to show a plot of $t^*$ from the mean-field calculation, compared with $\tau_{\textmd{active}}$ obtained numerically in the SI}.

In attractive colloids, the long-time behavior of MSD's has been used to analyze the gel-glass transition~\cite{C4SM00199K}.  In the next section, we quantify the dynamical arrest of the active states using this measure.    The phase diagram shown in Fig. \ref{fig1} (g) is constructed from these measurements.
The states with percolated regions of immobile particles, seen in Figs.~\ref{fig1} (e)-(f), are dynamically arrested gel-like states~\cite{Cates_2004}. The non-arrested states (Fig. \ref{fig1} (d)) resemble fluids with suspended ``beads'' of gels~\cite{Cates_2004}.  
In the arrested states, the long time behavior of $Q(t)$, which is the analog of the zero-wavevector, four-point susceptibility, $\chi_4 (q=0,t)$~\cite{PhysRevE.96.042605}, resembles that observed at small wave vectors in a model of chemical gelation~\cite{Candia2017}.  A striking feature of the arrested and non-arrested states at high activities is the appearance of ``voids''. We show below that the appearance of these voids is a manifestation of an extreme form of MIPS for these non-rotating particles that can be understood from ASEP dynamics. Physical gels arise from arrested phase separation~\cite{Zaccarelli_2007}, the gel-like states in our active lattice gas similarly seem to arise from arrested MIPS.

%%%%%%%%%%%%%%%%%%%%%%%%%%%%%%%%%%%%%%%%%%%%%%%%%%%%%%%%%%%%%%%%%%%%%%%
\section{ Collective Arrest, presence of ``Absorbing States"}
\label{sec:MSD}
In this section, we explore the appearance of arrested states through measurements of the MSD of individual particles.
The MSD of all particles ($i = 1,...,N$) in the system is defined as 
\begin{equation}
R^2(t) = \frac{1}{N}\sum_{i=1}^{i=N} \left| \vec{r}_i(t) - \vec{r}_i(0) \right|^2.
\label{msd_equation}
\end{equation}
In order to determine if states are arrested, we study the log derivative of this MSD for each system, defined as
\begin{equation}
\gamma(t) = \frac{d \log R^2(t)}{d \log t},
\label{gamma_definition}
\end{equation}
which has been used for studying dynamics in a model of gelation~\cite{C4SM00199K}.   
This observable is an instantaneous exponent that indicates whether the MSD is diffusive ($\gamma = 1 $), ballistic ($\gamma = 2$), or arrested ($\gamma = 0$).  
If there is percolation of a dynamically arrested  phase, then $\gamma (t) \rightarrow 0$ at long times.   Since the percolated phase acts as a solid boundary for the fluid-like particles, this is expected:  for a random walk confined within a box, it is well known that the MSD crosses over from diffusive growth $R^2 \sim t$ to a flat plateau $R^2 \sim t^{0}$ once the walker reaches the walls.

\begin{figure}[t!]
\includegraphics[width=0.5\textwidth]{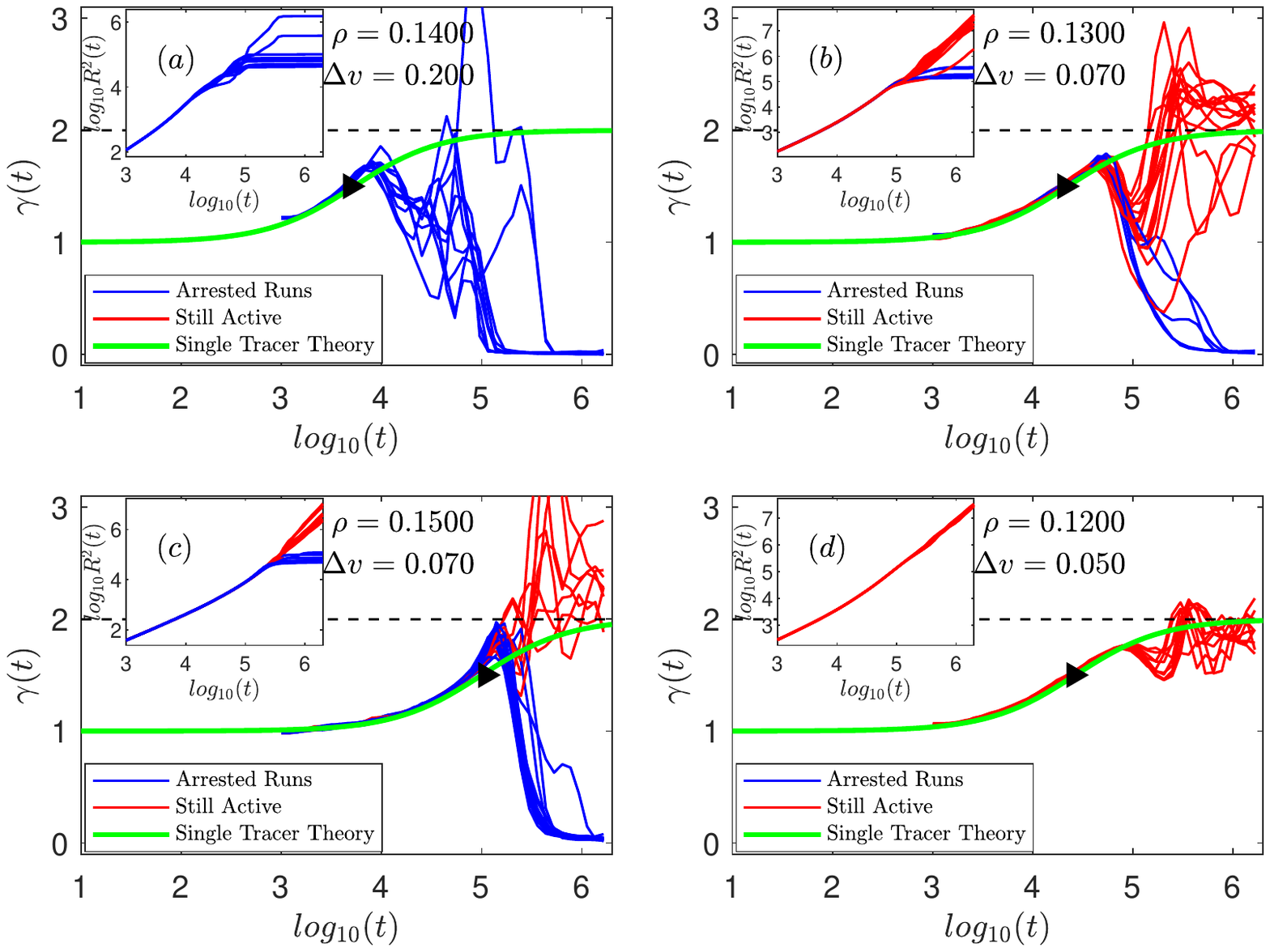}
\caption{Mean-squared displacements (MSD) $R^2(t)$ (insets) and their logarithmic derivatives $ \gamma(t) $ (main panels) at (a) $\rho = 0.14, \Delta v = 0.2 $, (b) $ \rho = 0.13, \Delta v = 0.07$, (c) $\rho = 0.15, \Delta v = 0.07$, and (d) $\rho = 0.12, \Delta v = 0.05$. Blue curves show ``arrested" states ($\gamma < 0.5$), while red curves show non-arrested states. The ensemble average $ \langle \gamma(t) \rangle $ is shown in black, and the mean-field prediction for a single active tracer is shown in green. The single tracer prediction crosses over from diffusive  ($\gamma= 1$) to ballistic ($\gamma= 2$) at a characteristic time $t^*$ (black arrows), which we estimate from $\gamma(t^*) = 3/2$.}
\label{fig2}
\end{figure}

Fig. \ref{fig2} shows MSD and $ \gamma(t)$ measurements for individual runs at large and intermediate activity values, $ \Delta v = 0.20 $ and $ \Delta v = 0.07 $. 
We can construct a mean-field model for the behavior of $ \gamma(t) $ in the non-arrested states  by using known results about the dynamics of a single non-rotating active tracer moving in a background of passive particles with density $ \rho $~\cite{chatterjee2019motion}. 
%In the limit of weak activity, $\Delta v =  v_+ - v_0 \rightarrow 0 $, the mobility of the tracer is equal to the diffusion coefficient of the passive particles $ v_p(\rho) = D(\rho) \Delta v $. 
For a non-rotating active tracer moving on a lattice at $\rho = 0$, the MSD is given by $\Delta r_i ^2(t) = D_0 (4 + \Delta v) t + D_0^2 {\Delta v}^2 t^2$~\cite{whitelam2018phase}.  This form may be generalized to higher densities by  introducing a density-dependent diffusion coefficient $D(\rho)$, giving 
\begin{equation}
\Delta r_i ^2(t) =  D(\rho) (4 + \Delta v) t + D(\rho)^2 \Delta v^2 t^2.
\end{equation}
Note that this requires a measurement of $D(\rho)$ for the passive lattice gas. For a single active tracer, therefore, $\gamma(t)$ displays a smooth crossover from $1$ to $2$ at a characteristic timescale, which we can estimate from $\gamma(t^*) = 3/2$, yielding $t^* = (4 + \Delta v)/({\Delta v}^2 D(\rho))$. This crossover time increases as the density is increased or activity is decreased. This  is the same trend as exhibited by $\tau_{\textmd{active}}$.   Thus, any non-trivial collective behavior arising from the activity appears once the ballistic motion takes over.

%Comparing this single particle theory to the collective MSD (Fig. \ref{fig2}) shows a very good agreement at short times. For longer times, there is a transition to new collective behaviour once 

%In Fig. \ref{fig2}(c) we construct a map of the probability of becoming arrested for different $(\rho,\Delta v)$  

We define arrested runs to be those for which $ \gamma(t_{\textrm{max}}) < 0.5$. For strong activities all runs fall into the arrested class, whereas at lower values of $\Delta v$ only a finite fraction of runs become arrested.  When considering the ensemble of possible dynamical trajectories, we observe two different types of activity-driven behavior: arrested states for which  $\gamma (t) \rightarrow 0$ at long times, and active states with $\gamma (t)$ fluctuating around a value of $2$.   It is clear from Fig. \ref{fig2} that the differentiation between arrested and non-arrested runs emerges only at times longer than $t^*$.

%{\color{red} CM: this summary of results in \cite{C4SM00199K} is inaccurate. 
The  behavior of $\gamma (t)$  offers the  clearest contrast between activity-induced arrest, as  seen in our model, and the attraction-induced gelation seen in passive colloids. In passive colloids, diffusing particles become arrested either to indicate a glass or gel transition. In contrast,  as seen from Fig. \ref{fig2},  it is the persistent motion, indicated by the ballistic behavior, that leads to arrest in the active hard crosses. The coarse-grained model for density inhomogeneity that we present in the next section is consistent with this picture.  

In passive systems, low-density gels can exhibit sub-diffusive behavior at intermediate times but $\gamma(t)$ asymptotes to unity since there are always particles that have finite mobility and can diffuse~\cite{C4SM00199K}. In fact, the MSD exhibits only weak signatures of gelation in passive colloids~\cite{Zaccarelli_2007,PhysRevE.81.040502,PhysRevLett.98.135503}. The behavior we observe is akin to colloids trapped in a porous environment~\cite{Kurzidim2011}. Since the arrested states (Figs. \ref{fig1} (e)-(f)) are characterized by a percolating network of immobile particles with no gaps between them, all the active crosses are effectively trapped and thus $\gamma(t) \rightarrow 0$ at long times. A similar suppression of diffusion was demonstrated in the 2D kinetically constrained model studied in \cite{ghosh2014jamming}, and also in the context of diffusion through biological anisotropic fibrous environments~\cite{gomez2019}. In the non-arrested states (Fig. \ref{fig1} (d)), there are pores through which the particles can escape and thus they exhibit sub-diffusive behavior at intermediate times, as seen in Fig. \ref{fig2}, but $\gamma(t) \rightarrow 2$ at long times as the particles recover their persistent, ballistic motion. 
 
%For the non-arrested runs, $\gamma(t)$ asymptotically approaches the ballistic value $\gamma \approx 2$. 
The color map of  arrested and non-arrested states shown in Fig. \ref{fig1} (g) was constructed from a measurement  of the probability of becoming arrested at  different values of $(\rho,\Delta v)$.  Between $10$ to $35$ runs were conducted at each set of parameter values, and the color bar indicates the fraction of those runs for which $ \gamma(t_{\textrm{max}})<0.5$. Fig. \ref{fig1} (g) shows that larger activity is required to generate arrested states at lower densities. For states in the range $(\rho>0.16,\Delta v<0.03)$, none of the states become arrested within our simulation time since the time $t^*$ needed for the active particles to exhibit ballistic motion is dramatically slowed down by both the small value of the diffusion coefficient $D(\rho)$, as well as the very weak activity $\Delta v $. 
Consequently, the states in this range of densities resemble those found in the passive system. 
%Contour lines of the time scale $t^*$ show that the simulation time limit would need to be made ten times larger or more in order to see if states in this range eventually become arrested. 
 
%%%%%%%%%%%%%%%%%%%%%%%%%%%%%%%%%%%%%%%%%%%%%%%%%%%%%%%
%%%%%%%%%%%%%%%%%%%%%%%%%%%%%%%%%%%%%%%%%%%%%%%%%%%%%%%
%%%%%%%%%%%%%%%%%%%%%%%%%%%%%%%%%%%%%%%%%%%%%%%%%%%%%%%
%%%%%%%%%%%%%%%%%%%%%%%%%%%%%%%%%%%%%%%%%%%%%%%%%%%%%%% 

%%%%%%%%%%%%%%%%%%%%%%%%%%%%%%%%%%%%%%%%%%%%%%%%%%%%%%%
%%%%%%%%%%%%%%%%%%%%%%%%%%%%%%%%%%%%%%%%%%%%%%%%%%%%%%%
\begin{figure*}
  \includegraphics[width=1\textwidth]{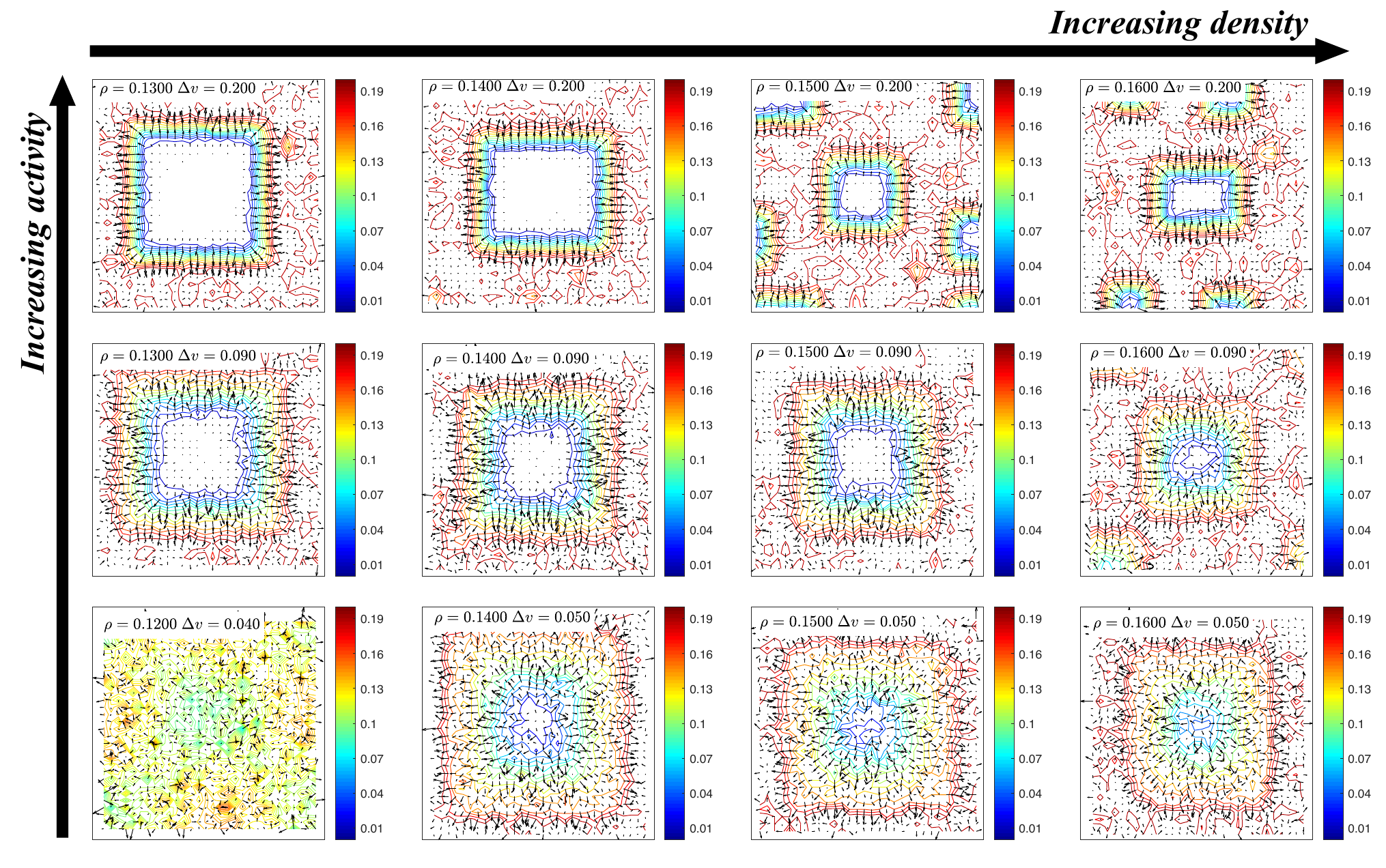} 
\caption{\label{2d_interface_samples} Contour maps of the coarse-grained density profile of arrested states at the longest simulation time $ t_{\textmd{max}} = 2 \times 10^6$. We used a coarse-graining box size with area  $L^2/900$, so that each box encloses $15^2$ lattice sites. The colorbar represents the local density which varies from $0 \le \rho \le 0.2$. The arrows represent the gradient of the density field, the lowest density point for each figure has been shifted to the center of the frame (making use of the periodic boundary conditions). Rows show fixed activity $\Delta v = 0.20, 0.09, 0.05$, and columns show densities increasing left to right $ \rho = 0.13, 0.14,0.15,0.16$. Density fluctuations for a typical non-arrested active liquid state  $(\rho=0.12,\Delta v = 0.040)$ are shown in the lower left corner. As the density is increased at a fixed activity, the width of the solid phase increases, and the active liquid fills the void region until it disappears. We observe that the width of the interface region is controlled solely by the activity $ \Delta v$.}
\end{figure*}
%%%%%%%%%%%%%%%%%%%%%%%%%%%%%%%%%%%%%%%%%%%%%%%%%%%%%%%
%%%%%%%%%%%%%%%%%%%%%%%%%%%%%%%%%%%%%%%%%%%%%%%%%%%%%%%

\section{coarse grained density profiles and hydrostatic lengthscale}
\label{sec:ASEP}
%asymmetric simple exclusion process 

Since dynamic differentiation emerges between different realizations of the simulations over a range of activity and density values, a further question arises about how these different classes of states differ structurally at long times. For our system of non-rotating, infinitely persistent active crosses, the morphology of the arrested states (see Figs. \ref{fig1} and \ref{2d_interface_samples}) depict voids  coexisting with 
a solid-like ($ \rho_{\textmd{solid}} \approx 0.19 $), highly immobile system spanning network, and a ``wetting'' layer of particles with intermediate mobility and density.
%appear to be an analog of MIPS in which the liquid phase has been replaced by a solid-like ($ \rho_{\textmd{solid}} \approx 0.19 $), highly immobile system spanning network.  
%Due to the persistent motion of particles along their oriented direction, the active particles accumulate on these solid surfaces enhancing the clustering that leads to solidification.  This process creates macroscopic ``voids'' separated from the solid by a fluid-like interface.  
The appearance of voids seem to be a feature of MIPS in the zero-rotation limit~\cite{mandal2019extreme,liao2018criticality}.

Starting from a random orientation of particles, as the dynamics of the system progresses, clusters of particles emerge with opposing active orientations, i.e. oriented towards each other, forming a solid-like high density region. At large times, mobile particles have an average active drift towards fluid-solid interfaces, as the particles oriented away from an interface have had sufficient time to travel to the other boundaries in the system. This active flux towards the interface causes an increasing density in the vicinity of the solid, giving rise to a diffusive current away from the solid. 
In order to model this process, we coarse grain the system   to construct a spatially varying density field $\rho(x,y)$. We consider a 1D section of the system perpendicular to an interface (oriented in the $y$-direction for convenience) between an active fluid and the solid, giving rise to a linear density profile $\rho(x)$. The exclusion due to particles in adjacent rows as well as their lateral diffusion give rise to correlations, which we ignore for large enough coarse graining blocks. This preference for biased motion perpendicular to the interface can also be modeled using the well-known ASEP model which incorporates both the hard-core exclusion along with diffusion and biased motion.  The steady state density profile in the arrested states can then be derived from a hydrodynamic treatment of  the ASEP ~\cite{lebowitz1988microscopic}, as we show below. 

The ASEP is a paradigmatic model where many exact statements can be made regarding the coarse grained dynamics in a non-equilibrium system. We can therefore, through this mapping, write equations for the coarse grained densities appearing at late times in our active crosses system. In steady state, the density $\rho(x)$ is independent of time, and there is no net particle current between the different coarse grained blocks. There are two components to this current determined by the density profile $\rho(x)$:
(1) A diffusive (or thermal) current $J_T$ arising due to the spatial variations in density, which to lowest order is
\begin{equation}
J_T = -D \frac{\partial \rho(x)}{\partial x},
\label{eq:diffcurrent}
\end{equation}
(2) An active current $J_A$ proportional to the density of particles. However, if neighboring blocks are at high densities, this current decreases due to exclusion. Once again to lowest order we have
\begin{equation}
J_A = \alpha \Delta v \rho(x) \left( \rho_{\textmd{solid}} - \rho(x) \right),
\label{eq:activecurrent}
\end{equation}
where $\alpha$ is an as yet undetermined proportionality constant.
These are essentially the mean field currents in ASEP  \cite{bodineau2006current}.

%%%%%%%%%%%%%%%%%%%%%%%%%%%%%%%%%%%%%%%%%%%%%%%%%%%%%%%
%%%%%%%%%%%%%%%%%%%%%%%%%%%%%%%%%%%%%%%%%%%%%%%%%%%%%%%
\begin{figure}[t!]
\includegraphics[width=0.5\textwidth]{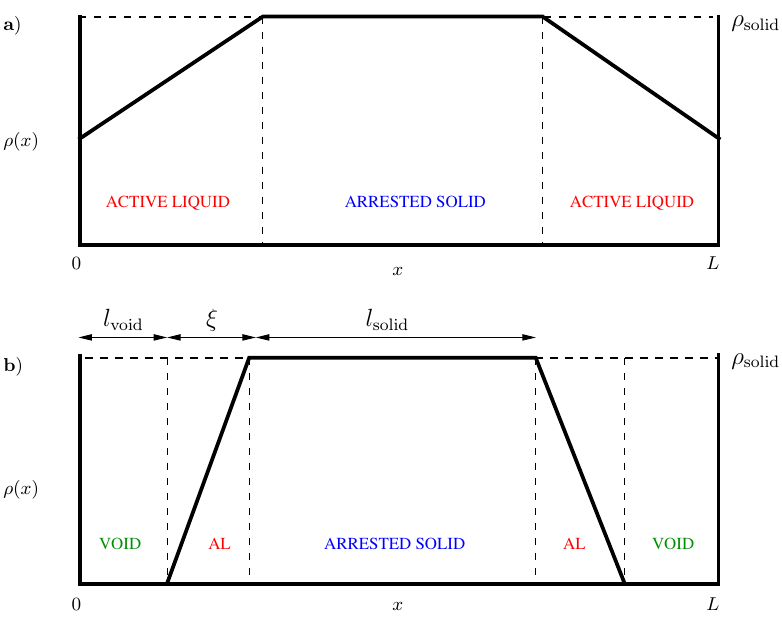}
\caption{One dimensional linear density profiles illustrating the non-equilibrium phase classification based on the observed configurations and local density distributions. (a) A dense solid region along with an active liquid interface which fills the remaining space available in the system (see, for example,  bottom row in Fig.~\ref{2d_interface_samples}). (b)  Solid region, bordered by a narrow active liquid (AL) interface of total width $ \xi $. The remaining area in the system is left empty of particles, creating a void (see, for example, top two rows in Fig.~\ref{2d_interface_samples}).}
\label{fig_linear_profiles}
\end{figure}
%%%%%%%%%%%%%%%%%%%%%%%%%%%%%%%%%%%%%%%%%%%%%%%%%%%%%%%
%%%%%%%%%%%%%%%%%%%%%%%%%%%%%%%%%%%%%%%%%%%%%%%%%%%%%%% 
 
\begin{figure}[t!]
\includegraphics[width=0.45\textwidth]{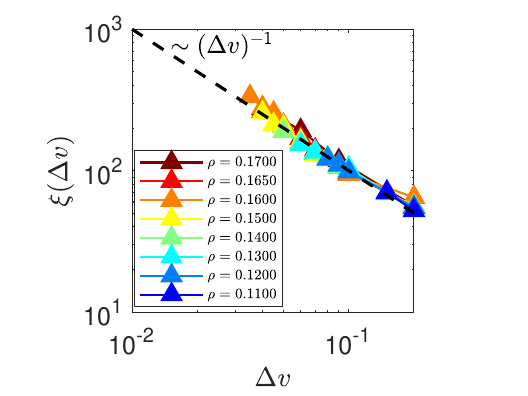}
\caption{The divergence of the lengthscale of the ``wetting active liquid" as the activity is decreased along with the theoretical prediction in Eq. (\ref{lengthscale_equation}). The data displays a good agreement with the $\Delta v^{-1}$ decay. The dashed black line shows the best fit $ \xi = \frac{10}{\Delta v} $.}
\label{active_width_figure}
\end{figure}

In steady state, the net currents are zero. Hence, combining Eqs. (\ref {eq:diffcurrent}) and (\ref{eq:activecurrent}), we obtain:
\begin{equation}
\frac{\partial \rho(x)}{\partial x} = \frac{\alpha \Delta v}{D} \rho(x) \left( \rho_{\textmd{solid}} - \rho(x) \right).
\label{eq:ASEP}
\end{equation}
The only two homogeneous solutions to this equation are voids with $\rho=0$ and the solid state with $\rho=\rho_{\textmd{solid}}$. Eq. (\ref{eq:ASEP})  is the logistic equation in space, which generates sigmoidal solutions.
The saturation values represent the solid and void regions, whereas the decaying part represents the wetting active fluid. At large distances this solution decays as $\exp[ - (\alpha  \Delta v / D) x ]$, implying a wetting lengthscale
\begin{equation}
\xi = \frac{D} { \alpha \Delta v}.
\label{lengthscale_equation}
\end{equation}
We note that this ``hydrostatic lengthscale" diverges in the limit of zero activity.
This divergence as the activity is decreased is shown in Fig.~\ref{active_width_figure} along with the theoretical prediction from Eq. (\ref{lengthscale_equation}) showing near perfect agreement with the $\Delta v^{-1}$ decay. 
Note that Eq. (\ref{lengthscale_equation}) does not involve the global density of the system, and the correlations between  rows  in our two-dimensional lattice can provide non-trivial corrections to the derived behavior for larger densities.

%%%%%%%%%%%%%%%%%%%%%%%%%%%%%%%%%%%%%%%%%%%%%%%%%%%%%%%
%%%%%%%%%%%%%%%%%%%%%%%%%%%%%%%%%%%%%%%%%%%%%%%%%%%%%%%

%%%%%%%%%%%%%%%%%%%%%%%%%%%%%%%%%%%%%%%%%%%%%%%%%%%%%%%

%%%%%%%%%%%%%%%%%%%%%%%%%%%%%%%%%%%%%%%%%%%%%%%%%%%%%%%
\section{Non-Equilibrium Phase Diagram}
\label{sec:density}

%We next construct a phase diagram using the observed forms of the density profiles in our system.  
%As shown in Fig \ref{fig_phase_examples}, three types of  density profiles are observed: an active, liquid-like phase with density fluctuations but no immobile solid regions, a coexistence of this ``active liquid" with solid regions, and finally at large activities, a solid network, punctuated by voids, and an active liquid interface separating the two.  By considering the conditions that must be satisfied at a given density and activity to create each of these configurations, we can derive phase boundaries between these three types of states. 
As seen in Fig.~\ref{2d_interface_samples}, two types of profiles are observed in the arrested states: coexistence of an ``active liquid" with solid regions (bottom row), and  a solid network, punctuated by voids that have a characteristic size, and an active liquid interface separating the two.  In addition, there are non-arrested, active liquid states with density fluctuations of amplitude much smaller than the solid density.  In the previous section, we showed that our theory correctly predicts the variation of the width of the interface.  In this section, we extend our analysis to construct a non-equilibrium phase diagram of the arrested states, and provide a theory for the emergent length scales characterizing the voids and the solid regions.

We can derive phase boundaries between the three types of states by considering the conditions that must be satisfied at a given density and activity to create each of these configurations. 
For the arrested states at strong activities, the dense immobile solid ($\rho_{\textmd{solid}} \approx 0.19$) is bordered by an active liquid interface that can be fit by a linear profile (as an approximation to the sigmoidal solutions of Eq. (\ref{eq:ASEP})) with  slope $ m = \frac{d \rho}{d x} $ and width $ \xi $, such that $ \xi m = (\rho_{\textrm{solid}} - \rho_{\textrm{void}}) = 0.19 $. 
%%
%As we show in Fig \ref{fig_phase_examples}, three kinds of non-equilibrium phases appear in this system: an active, liquid-like phase with density fluctuations but no immobile solid regions, a coexistence of this ``active liquid" with solid regions, and finally at large activities, a solid network, punctuated with pores containing a vacuum, and an active liquid interface separating the vacuum and the solid. We find that the solid regions always occur in conjunction with an active liquid interface of a fixed width $ \xi $ depending solely on the activity $\Delta v$ as predicted in Eq. (\ref{lengthscale_equation}). By consideration of the conditions that must be satisfied at a given density and activity to create such configuration, we can derive phase boundaries between these three kinds of states. 
We use these linear density profiles (as shown in Fig.~\ref{fig_linear_profiles}) for all states, including the active liquid, in the computation of the phase diagram. The total length of the system is fixed at $L$, and the total number of particles in the system is conserved as $N = \rho L$. 

The first condition needed to create a solid region along with an active liquid interface is that there must be enough mass available in the system to populate both these regions (as in Fig.~\ref{fig_linear_profiles} (a)). The total area under the trapezoid must conserve the total number of particles in system at a given global density $\rho$, yielding
\begin{equation}
N = \rho  L = (\rho_{\textrm{solid}}) (\xi   + l_{\textrm{solid}}).
\label{density_equation}
\end{equation}
Since the interface width is determined by the activity (Eq. (\ref{lengthscale_equation})), the active liquid region can only contain a fixed activity-dependent mass, and consequently the width of the solid region depends on both the density and the activity as 
\begin{equation}
l_{\textrm{solid}} = \frac{\rho}{\rho_{\textrm{solid}}}L - \frac{D}{\alpha \Delta v}.
\label{solid_width_equation}
\end{equation} 
The solid first appears at the point where $ l_{\textrm{solid}} = 0 $, yielding the equation for the solid-active liquid phase boundary 
\begin{equation}
\Delta v = \frac{D }{\alpha L} \frac{\rho_{\textrm{solid}}}{\rho}.
\label{phase_boundary_eq1}
\end{equation}
Below this activity value, the interface width is larger than $ L /2 $, therefore we expect the active liquid phase to contain all the mass in the system. 

%%%%%%%%%%%%%%%%%%%%%%%%%%%%%%%%%%%%%%%%%%%%%%%%%%%%%%%
%%%%%%%%%%%%%%%%%%%%%%%%%%%%%%%%%%%%%%%%%%%%%%%%%%%%%%%
\begin{figure}[t!]
\includegraphics[scale=0.45]{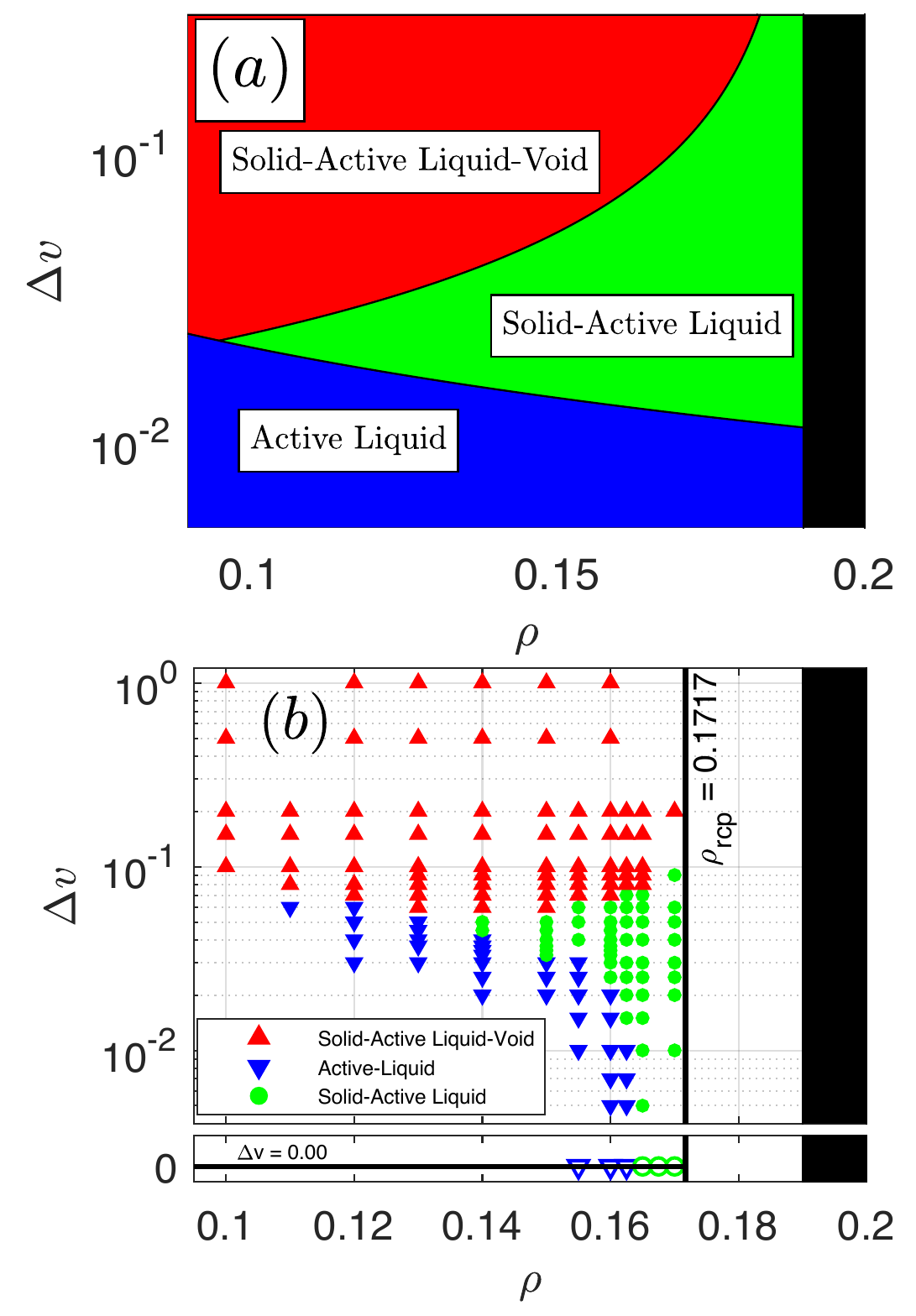}
\hfill
\caption{(a) Classifications of the activity-induced phases, based on linear density profiles. The solid lines indicate the theoretically predicted boundary between the active liquid and solid + active liquid regions in Eq. (\ref{phase_boundary_eq1}), and the activity beyond which void regions open up (predicted by Eq. (\ref{phase_boundary_eq2})). We have used the observed value $ \frac{D}{\alpha}=10 $, $ \rho_{\textrm{solid}} = 0.19 $, and the simulated system size $ L = 450$. (b) Phase diagram obtained from the numerically sampled    phase space of the active lattice gas, including the passive and the active states. Each point is classified as either pure active liquid, solid + active liquid, or solid + active liquid along with void regions. Open symbols denote the transition from the passive liquid to the aging glassy regime. These classifications are based on ensemble-average density profiles. The topology of this phase diagram is  the same as that observed in attractive colloids~\cite{Zaccarelli_2007,Cates_2004,Royall:2018aa}. The black region denotes densities larger than $\rho_{\textrm{solid}} = 0.19$. Note that the predicted phase diagram shown in (a) is valid only at finite activities and is expected to accurately capture the physics only at large enough activities where activity-induced correlations overwhelm  the correlations intrinsic to the passive hard crosses at $\rho \approx \rho_{rcp}$.  This is because, as discussed in the text, the  coarse-grained model  includes only simple exclusion, not the extended N3 exclusion.}
%Significant differences between (b) and (a) are (i)  the appearance of a liquid-solid coexistence regime in the passive limit, and (ii) the absence of any effect of $\rho_{\textrm{rcp}}$ in (a). This is because, as discussed in the text, the ASEP-based coarse-grained model does not capture the physics of the passive hard crosses.}
\label{fig_nonequilibrium_phase_diagram}
\end{figure}
%%%%%%%%%%%%%%%%%%%%%%%%%%%%%%%%%%%%%%%%%%%%%%%%%%%%%%%
%%%%%%%%%%%%%%%%%%%%%%%%%%%%%%%%%%%%%%%%%%%%%%%%%%%%%%%

So far we have only imposed conservation of mass on the system. To determine when voids first appear, we must also consider whether the shape of the trapezoidal solid-active liquid profile can fit within the total space of the system.  Since none of the regions overlap, their lengths sum to the system size $ L $. We therefore have (see Fig. \ref{fig_linear_profiles} (b)) 
\begin{equation}
L = l_{\textrm{solid}}+l_{\textrm{void}} + 2\xi.
\label{length_equation}
\end{equation}
Since the thickness of the solid region is constrained by conservation of mass, Eq. (\ref{solid_width_equation}), we can solve for how much space remains available for the void region, 
\begin{equation}
l_{\textrm{void}}= \left(1-\frac{\rho}{\rho_{\textrm{solid}}} \right)L - \frac{D}{\alpha \Delta v}.
\label{void_length_eq}
\end{equation}
To determine when a void region may first appear, we find the point $l_{\textmd{void}} = 0$.
This yields the second phase boundary between states containing voids and those without voids
\begin{equation}
\Delta v = \frac{D}{\alpha L}\frac{1}{1-\rho/\rho_{\textrm{solid}}}.
\label{phase_boundary_eq2}
\end{equation}
Below this activity the active liquid is confined within a space smaller than its preferred width $\xi$. 

Phase boundaries based on the 1D linear profile analysis, presented above, are shown in Fig.~\ref{fig_nonequilibrium_phase_diagram} (a). In Fig.~\ref{fig_nonequilibrium_phase_diagram} (b) we show that a numerical classification of the three types of non-equilibrium phases agrees qualitatively with the 1D theory.  It is straightforward to extend the treatment developed in this section to a derivation of the phase boundaries based on 2D density profiles. We find that the qualitative features of the resulting phase diagram do not change as compared to the 1D case studied here.
The states in the phase identified as the solid-active liquid, based on the density profiles, are dynamically the aging glassy states shown in Fig. \ref{fig1}.  The arrested, gel-like states, shown in Fig. \ref{fig1} are in the solid-active liquid-void region of Fig. \ref{fig_nonequilibrium_phase_diagram}.

%%%%%%%%%%%%%%%%%%%%%%%%%%%%%%%%%%%%%

%%%%%%%%%%%%%%%%%%%%%%%%%%%%%%%%%%%%%

In Appendix \ref{appendix_d} we show that the variation of the phase boundaries  with system  size is  consistent with the predictions of the theory derived above.
%More physical inputs are needed to describe the behaviour of the phase boundary near the glass transition.
In the infinite system size limit, the ASEP analysis implies that voids can  be accommodated at any density and activity, as the phase boundaries become lines with infinite slopes at $\rho = 0$ and $\rho=\rho_{\textmd{solid}}$.  This feature is a consequence of the simple exclusion process, which cannot describe the non-trivial correlations in the passive, hard-cross system (c. f. Fig. \ref{fig_nonequilibrium_phase_diagram} (b)).  We are currently exploring avenues for incorporating these effects, possibly through the construction of a large-deviation function that incorporates the physics in  of void-solid coexistence of persistent hard crosses, encapsulated in the ASEP model, and the glass transition physics of passive hard crosses.

%%%%%%%%%%%%%%%%%%%%%%%%%%%%%%%%%%%%%%%%%%%%%%%%%%%%%%%
%%%%%%%%%%%%%%%%%%%%%%%%%%%%%%%%%%%%%%%%%%%%%%%%%%%%%%%
\begin{figure}[t!]
\includegraphics[width=0.5\textwidth]{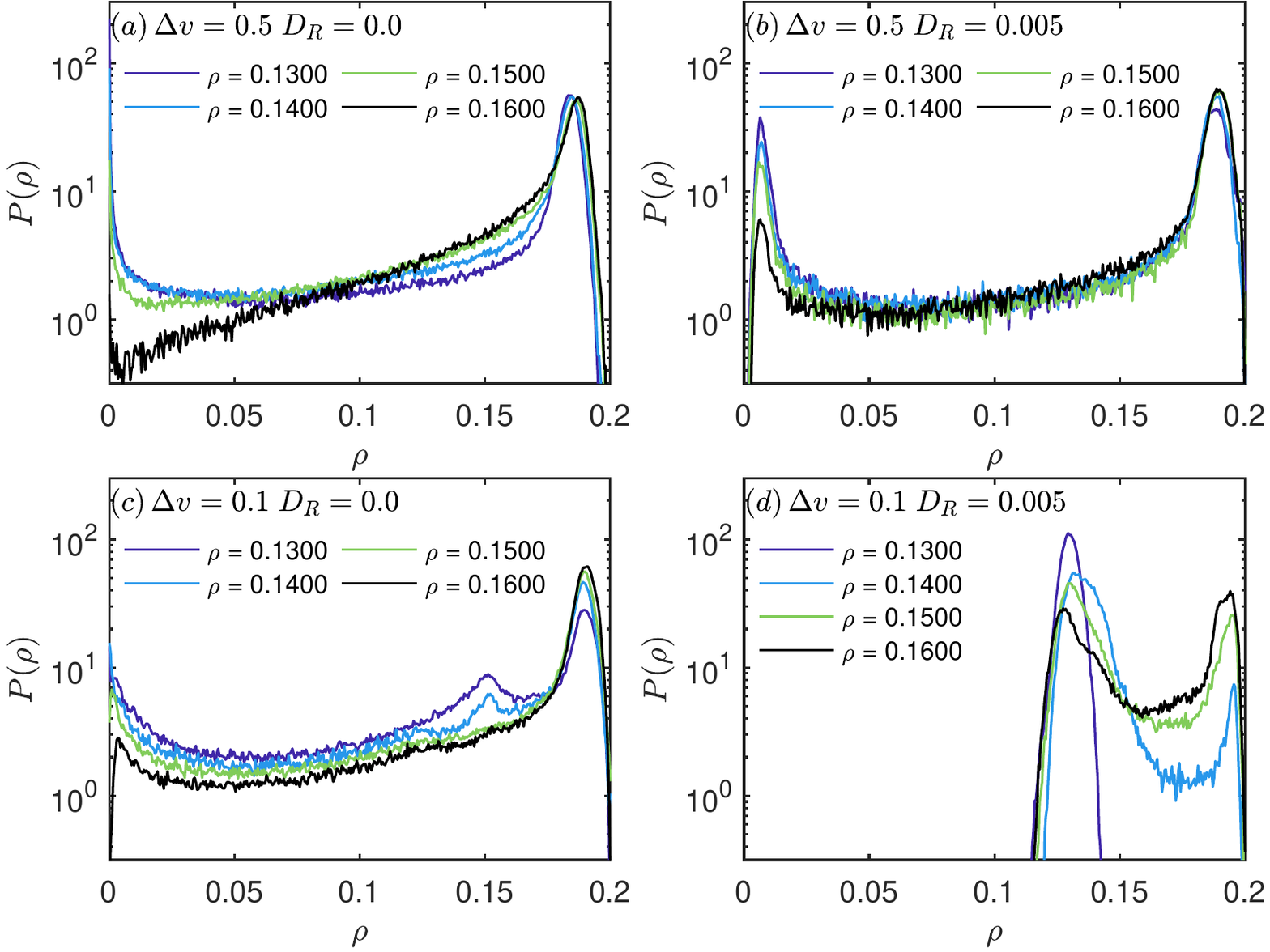}
\caption{Comparison of the local density distributions for persistent ($ D_R = 0 $) and finite rotation rate ($ D_R = 0.005 $) active hard crosses. At $ \Delta v = 0.5 $, the zero density peak corresponding to the empty void regions at $D_R=0$ (a) is replaced by a well-defined low density peak at $D_R=0.005$ (b). For the smaller activity value, the non-rotating crosses show phase separation into solid and void-like regions, with an intervening active liquid region (c). In contrast, at $D_R=0.005$,  there is a  simpler fluid-solid coexistence indicated by a clearly bimodal distribution (d).}
\label{density_dist_finite_and_zero_rot}
\end{figure}
%%%%%%%%%%%%%%%%%%%%%%%%%%%%%%%%%%%%%%%%%%%%%%%%%%%%%%%
%%%%%%%%%%%%%%%%%%%%%%%%%%%%%%%%%%%%%%%%%%%%%%%%%%%%%%% 

%%%%%%%%%%%%%%%%%%%%%%%%%%%%%%%%%%%%%%%%%%%%%%%%%%%%%%%%%%%%%%%%%%%%%%%
%%%%%%%%%%%%%%%%%%%%%%%%%%%%%%%%%%%%%%%%%%%%%%%%%%%%%%%%%%%%%%%%%%%%%%%

\section{Effect of Finite Rotation Rate}
\label{sec:finite_rotation}
%\label{appendix_c}

In this section we discuss the effect of a finite but small rotation rate on the arrested phase separation observed for non-rotating hard crosses ($D_R=0$) derived above.
%Given the presence of rotational locking among neighboring crosses, we expect the  global dynamical, gel-like arrest to persist even in systems with some amount of rotational relaxation. 
 We perform simulations in two regimes $D_R = 0.005, \Delta v = 0.1$, and $D_R = 0.005, \Delta v = 0.5$.  It should be noted that $D_R$ is the {\it attempted} rotation rate.  The non-convex shape of the hard-crosses leads to strong rotational locking that leads to a smaller effective rotation rate, which also decreases as the density increases.
%
%Lastly, we have investigated the dependence of the phase behaviors and arrested structures observed in the infinite persistence-time limit primarily studied in this work when this limit is relaxed and a small, finite rotation rate $ D_R = 0.005$

We find that at large activities $\Delta v \geq  0.5 $, a percolated, globally arrested solid network, similar to the ones observed at $D_R=0$,  is able to form in the same regime of densities as shown in Fig. \ref{fig1}. However, there are structural differences between the arrested states at zero and non-zero $D_R$.  Specifically,  the voids observed at $D_R=0$ are replaced by a low-density, gaseous fluid at $D_R=0.005$  as seen in Fig. \ref{finite_and_zero_rotation_comparison}.  The fluid phase still exhibits an inhomogeneous density profile with the highest densities occurring at the interface with the solid.  In addition,  a finite rotation rate seems to create a  rounding of the facets separating the solid from the fluid.  At lower activities, the fluid density becomes homogeneous, as seen in Fig. \ref{finite_and_zero_rotation_comparison2}.  The morphology of these states resemble those observed in the active-aging regime (Fig. \ref{fig1} (c)) rather than the  percolating arrested solid observed at $D_R=0$, seen in Fig. \ref{finite_and_zero_rotation_comparison2} (a).

%%%%%%%%%%%%%%%%%%%%%%%%%%%%%%%%%%%%%%%%%%%%%%%%%%%%%%%
%%%%%%%%%%%%%%%%%%%%%%%%%%%%%%%%%%%%%%%%%%%%%%%%%%%%%%%
\begin{figure}[t!]
\centering
\includegraphics{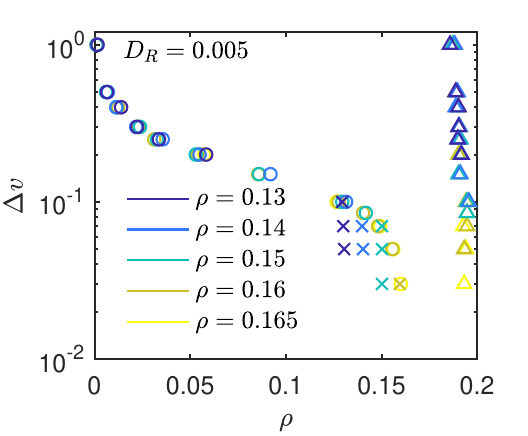}
\caption{\label{phase_diagram_finite_rot} Phase diagram at $D_R=0.005$ showing the binodal constructed from the peaks in ensemble density distributions with triangles marking the high-density peak and circles the low-density one. Crosses indicate  phase space points $ (\Delta v, \rho)$ at which  the system is in a homogeneous fluid state with a single peak at the global density in the density distribution. Colors indicate the global density $\rho$.   Density distributions were averaged over $5$ runs for each sampled phase space point.}
\end{figure}
%%%%%%%%%%%%%%%%%%%%%%%%%%%%%%%%%%%%%%%%%%%%%%%%%%%%%%%
%%%%%%%%%%%%%%%%%%%%%%%%%%%%%%%%%%%%%%%%%%%%%%%%%%%%%%%

The  ensemble-averaged density distributions shown in Fig. \ref{density_dist_finite_and_zero_rot}, demonstrate that at $D_R=0.005, $ the ``void'' peak disappears.  In addition, there is a  clear signature of two-phase coexistence between a fluid and a solid, with little dependence of the peak positions on the global density~\cite{klamser2018thermodynamic}.  With increasing activity, the peak marking the fluid density shifts significantly to lower densities whereas the solid peak stays roughly pinned $\rho \approx 0.19$.  Since the ensemble-averaged density distributions do not depend on the global density at a fixed activity (Fig. \ref{finite_and_zero_rotation_comparison2}), a binodal line can be constructed from the positions of the peaks, as shown in  Fig.~\ref{phase_diagram_finite_rot}.  
 Interestingly, this phase diagram looks  different from that of spherical ABPs~\cite{klamser2018thermodynamic}, and remarkably similar to the one observed in  
%This is in contrast to simulations of spherical ABPs in continuum where the fluid-solid coexistence and the gas-liquid MIPS region appear to be disconnected~\cite{klamser2018thermodynamic}. 
actively-driven dumbbell shaped particles, which are also non-convex~\cite{cugliandolo2017phase}.

We can analyze the stability of the $D_R = 0$ phase diagram to small but non-zero $D_R$. On the lattice, since the particles can only have discrete orientations, we can compute the mean free path of the particles in the persistent direction. Since the density of rotation events in time have a gap distribution $D_R \exp(-D_R t)$, and the average time between succesful rotation events scales $1/D_R$ \cite{chatterjee2019motion}, the average length traveled without a rotation move, is given by $\xi_R \sim (\Delta v) (1/D_R)$. In order for the rotational diffusion to affect the motion of the particles, there must be a sufficient time for the particles to move before reaching the dense regions and become a part of the wetting fluid.  Therefore, when the average length traveled in the persistence direction is comparable to the size of the voids $l_{\textrm{void}}$, the phases appearing will no longer correspond to the zero rotation limit. This provides a crossover value $D_R^{*} = \Delta v/ l_{\textrm{void}}$, above which the rotational diffusion takes on significant effect.  Since the size of the voids can be estimated from Eq. (\ref{void_length_eq}), we can estimate the crossover rotation rate below which voids are expected to appear in the system
\begin{equation}
\frac{1}{D_R^{*}} =\left(1-\frac{\rho}{\rho_{\textrm{solid}}} \right) \frac{L}{\Delta v} - \frac{D}{\alpha \Delta v^2}.
\end{equation}

For the values of $\Delta v$ used in the simulations,  and using the values $D/\alpha = 10$ and $L = 450$ deduced from the data,  we estimate the crossover value for density $\rho = 0.1$ to be $D_R = 0.005$ for $\Delta v = 0.1$, and  $D_R = 0.009$ for $\Delta v = 0.5$. Since the crossover value of $D_R$ decreases with $\Delta v$, we expect that as the activity is decreased, the qualitative features change from those observed in arrested states at zero rotation, and begin to resemble a MIPS  between a low and high density fluid, as is observed in Fig. \ref{finite_and_zero_rotation_comparison2}.

To summarize, we find that the most significant difference between hard crosses with $D_R=0$ and $D_R=0.005$ is  the disappearance of voids, which also leads to  changes in the phase diagram.   In earlier studies, the appearance of voids  has been associated with  active particles with  infinitely persistent self-propulsion directions~\cite{liao2018criticality}.  
%We expect that the rotational locking feature of hard crosses would lead to corrections to the crossover value of rotation rates for which voids appear in the system.  We plan to investigate the effect of rotational locking on the appearance of voids, and phase separation in more detail in the near future.
%An interfacial (wetting) fluid still separates the very low density, gaseous region from the arrested solid.   

\section{Discussion}

In this paper we have studied a lattice gas of active particles with a non-convex shape that leads to strong inhibition of rotations. We showed that unlike the usual MIPS observed in active particles with large rotational diffusion, highly persistent particles are able to arrest the phase separation leading to states that are characterized structurally by a void-solid coexistence with a  fluid of relatively  mobile particles  ``wetting'' the void-solid interface.  The voids have a characteristic size that depends on density and activity in contrast to coarsening, as they would if the phase separation was not arrested.  Dynamically,  the glassy dynamics characterizing the aging regime of the passive system transitions to complete arrest in the phase-separated regime.  In addition, activity enlarges the density range of the aging regime.  We also show that adding moderate rates of rotational diffusion does not change this picture qualitatively.  

By appealing to the dynamics of a single active tracer in a passive background of hard crosses~\cite{chatterjee2019motion}, we showed that the dynamic differentiation between arrested and non-arrested states appears at times longer than that needed for the tracer dynamics to  cross over from diffusive to ballistic.  Thus, the gel-like arrest, in contrast to the glassy caging dynamics of the passive system, is driven by activity. The morphology of the arrested states, however, closely resembles gelation or arrested phase separation in passive colloids with attractive interactions~\cite{Cates_2004,Royall:2018aa}. Our analysis shows a clear crossover from the passive glassy dynamics to gel-like arrest at high densities~\cite{Royall:2018aa}.  

We used the fact that our persistently active dynamics effectively causes one-dimensional motion against an interface, to map the late-time dynamics in the arrested states onto the Asymmetric Exclusion Process (ASEP). This microscopic mapping of the long time behavior of the system to a well-known lattice model allowed us to invoke well-established coarse graining procedures which we used to describe the non-trivial collective behaviour observed in our system. Building on this understanding, we used the predicted length scale to map out a non-equilibrium phase diagram that predicts non-trivial phases, a non-trivial topology, as well as non-trivial finite size scaling. While the position of the phase boundaries are not in quantitative agreement with our numerical results, the topology matches the numerical results over most of the phase space. The ASEP-based predictions fail at low activity and high densities because it takes into account only simple exclusion, and thus does not incorporate the shape-induced frustration and glassy dynamics in the passive system at high enough densities. 
%However, the non-equilibrium phase diagram based on these density profiles provides a qualitatively correct description of the phase boundaries at large activities.
%
%
% {\color{red} {\color{blue} CM: Safer to leave out this statemnent???}\sout{The non-equilibrium phase diagram obtained from ASEP captures the numerically determined phase diagram semi-quantitatively, therefore, we deduce that the underlying lattice in our model does not affect the behavior at long length scales and large time scales.} }The ASEP mechanism also provides a natural explanation for the length scales emerging in the morphology of the arrested states. 
%At low densities, voids coexist with an arrested solid:  activity drives  phase separation and arrest in the infinitely  persistent limit. 
% In passive colloids, gelation at low densities leads to strong density heterogeneities with void-like regions, however, at high densities the gels are significantly more homogeneous~\cite{Royall:2018aa}.  In contrast, the activity-driven gelation observed in persistent hard crosses displays void-solid coexistence even at high densities if the activity is high enough.   
%The hydrodynamic theory that we have presented offers a clear direction to pursue in understanding dynamical arrest in  continuum models of active colloids.  

In this paper we make concrete predictions about the phase behavior and the morphology of  arrested states when rotation of the active direction is strongly inhibited. This should occur naturally in collections of active entities with non-convex shapes~\cite{Wittkowski_Lowen}. Numerical simulations have explored MIPS~\cite{cugliandolo2017phase} and glassy dynamics~\cite{PhysRevE.96.042605} in active dumbbells. In these studies, the mechanism of de-correlation of the active direction is, however, related to thermal diffusion~\cite{PhysRevE.96.042605} and not independently controlled. It would be interesting to explore dynamical arrest in simulations of rigid non-convex shapes where the rotation rate is affected by rotational locking.   Experimental investigation of phase separation and arrest in collections of  active colloidal particles with non-convex shapes offer possibilities of testing our theoretical predictions.   We are not aware of any such experimental investigation, however, extending studies such as the Brownian dynamics of hard crosses~\cite{Zhao_2014} to include self propulsion seem feasible. 

%The phenomenon of MIPS has been convincingly demonstrated to be a consequence of the self-propulsion leading to an  effective attraction between ABPs~\cite{cates2015motility}.   The dynamical arrest discussed in this paper shows  that this activity-driven attraction can lead to physics that closely resembles gelation or arrested phase separation in passive colloids with attractive interactions~\cite{Cates_2004,Royall:2018aa}.   Our analysis shows a clear crossover from the passive glassy dynamics to gel-like arrest at high densities~\cite{Royall:2018aa}.  At low densities, voids coexist with an arrested solid:  activity drives  phase separation and arrest in this persistent limit.  In passive colloids, gelation at low densities leads to strong density heterogeneities with void-like regions, however, at high densities the gels are significantly more homogeneous~\cite{Royall:2018aa}.  In contrast, the activity-driven gelation observed in persistent hard crosses displays void-solid coexistence even at high densities if the activity is high enough.   An interesting direction for future research would be to understand the behaviour of particles in continuum based on the techniques developed in this paper.
%

\vspace{0.5cm}
%%%%%%%%%%%%%%%%%%%%%%%%%%%%%%%%%%%%%%%%%%%%%%%%%%%%%%%%%%%%%%%%%%%%%%%
%%%%%%%%%%%%%%%%%%%%%%%%%%%%%%%%%%%%%%%%%%%%%%%%%%%%%%%%%%%%%%%%%%%%%%%
\section*{Acknowledgments}
We thank Mustansir Barma, Pinaki Chaudhuri, Chandan Dasgupta, Madan Rao, Abhishek Dhar, Eli Eisenberg and Andras Libal for helpful discussions.
The work of CM and BC has been supported by the Brandeis MRSEC (NSF-DMR 1420382). CM was hosted as graduate fellow at KITP, with support from the National Science Foundation under Grant No. NSF PHY-1748958 and the Heising-Simons Foundation. This work was also partially supported by the Israel Science Foundation Grant No. 968/16 and by a grant from the United States-Israel Binational Science Foundation.  

\vspace{0.5cm}

\begin{appendix}

%\section*{\large Supplemental Material for\\ ``Dynamical Arrest of Persistently Active Hard Crosses"}

%In this document we provide supplemental figures and details of the calculations presented in the main text.

\twocolumngrid

%%%%%%%%%%%%%%%%%%%%%%%%%%%%%%%%%%%%%%%%%%%%%%%%%%%%%%%%%%%%%%%%%%%
%%%%%%%%%%%%%%%%%%%%%%%%%%%%%%%%%%%%%%%%%%%%%%%%%%%%%%%%%%%%%%%%%%%

\vspace{0.5cm}

\section{Dynamics of Passive Hard Crosses}
\label{appendix_a}

\begin{figure}[t!]
\centering
\includegraphics[width= 0.9\linewidth]{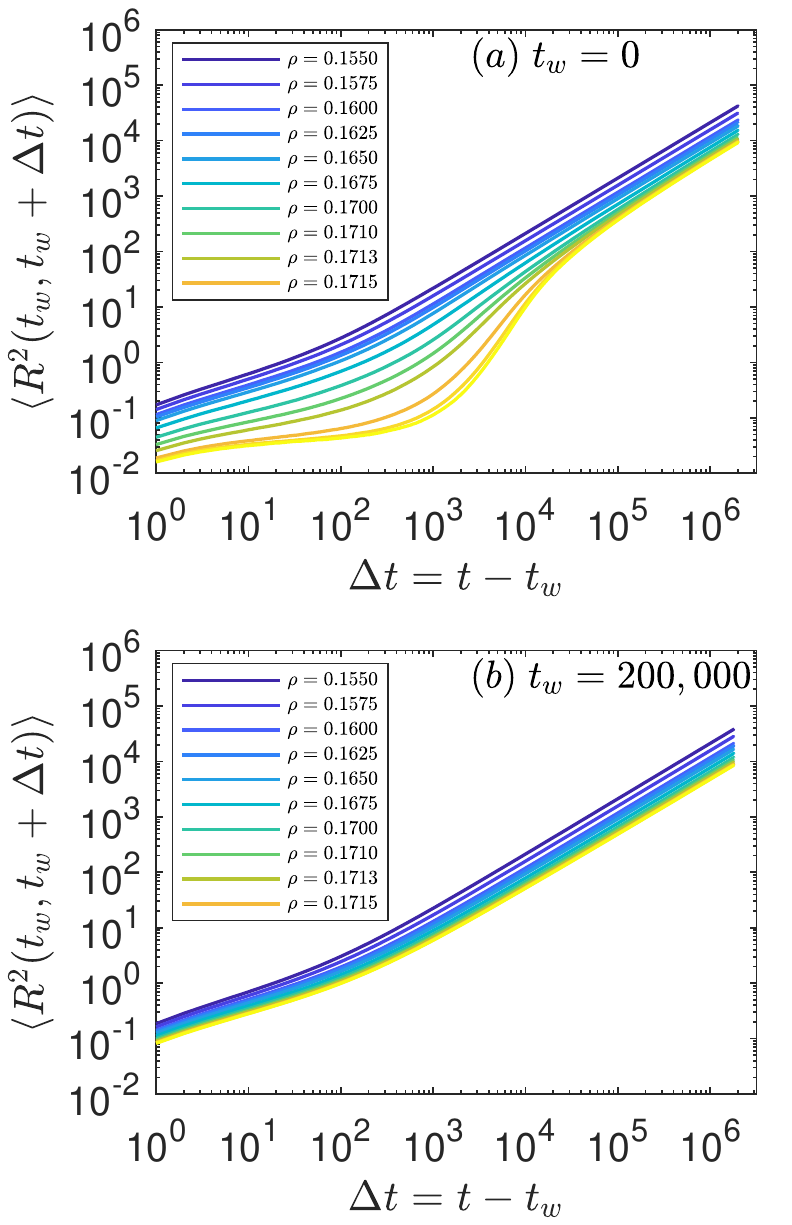}
\caption{Mean-squared displacements (MSD) of passive hard crosses starting from: (a) the  initial state produced by the RSAD protocol,  $ t_w = 0 $, and (b) $ t_w = 200,000$, showing significant dependence on $t_w$,  for densities $ \rho \ge 0.1650 $:  note the absence of the plateau observed  in (a) at  longer waiting times (b).}
\label{passive_MSD}
\end{figure}

\begin{figure}[h!]
\centering
\includegraphics[width= 0.9\linewidth]{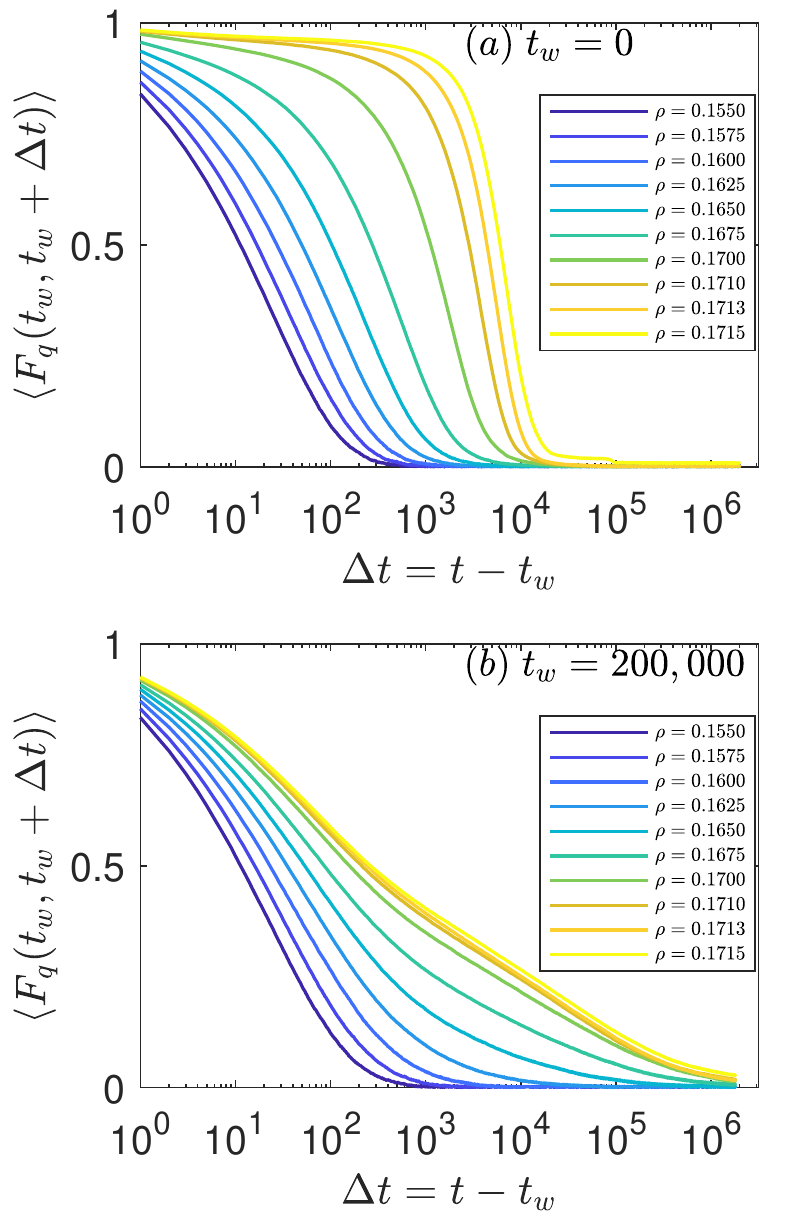}
\caption{Self-intermediate scattering functions (ISF) of passive hard crosses starting from:  (a) the  initial state produced by the RSAD,  $ t_w = 0 $, and (b) $ t_w = 200,000$, showing significant dependence on  $t_w$,  for densities $ \rho \ge 0.1650 $:  note the absence of the plateau observed  in (a) at  longer waiting times (b).}
\label{passive_ISF}
\end{figure}

\begin{figure}[t!]
\centering
\includegraphics[width= 0.9\linewidth]{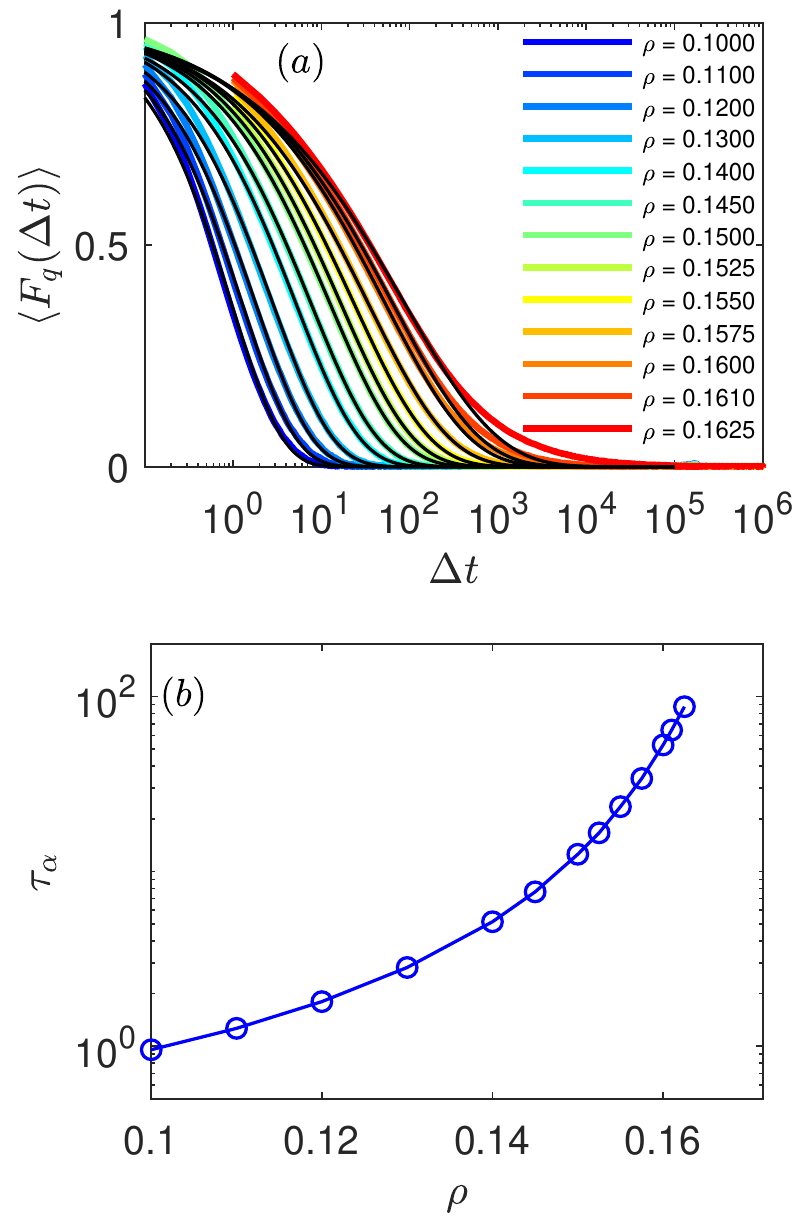}
\caption{(a) Stretched exponential fits (black lines) to the ISFs at densities $ \rho \le 0.1625 $, showing that the fit fails at $\rho=0.1625$.  The exponent,  $ \beta $,  decreases  from  $ 0.8 $ at $\rho=0.1$ to and $ 0.4 $ at $\rho=0.16$. (b) $ \tau_{\alpha} $ extracted from the stretched exponential fits.}
\label{ISF_stretched_fits}
\end{figure}

In this appendix, we provide a brief description of our analysis of slow dynamics and the ``glass transition'' in the passive hard cross system ($\Delta v= 0$)~\cite{rotman2009ideal,rotman2010direct,eisenberg2000random}.  This analysis was performed to establish a baseline for the dynamics in the absence of activity.  The initial states were created through a quench into the two-phase coexistence region using the Random Sequential Adsorption and Diffusion (RSAD) protocol~\cite{eisenberg1998random}.    The hard-crosses were then evolved according to the dynamics described in Section \ref{sec:model} in the main text, and two standard measures were used to probe the glassy dynamics ~\cite{berthier2011theoretical,debenedetti2001supercooled}: (i) the self-intermediate scattering function (ISF) and the (ii) mean-squared-displacement (MSD) of individual crosses.  The slowest relaxation occurs at the wave-vector with magnitude $q$ at which the static structure factor has a peak.  
{
For $N$ particles, the ISF  is defined as
\begin{equation}
F_{q}(t_w,t_w+\Delta t) =  \frac{1}{N} \sum_{j=1}^N \exp(i \vec{q} \cdot (\vec{r}_j (t_w+\Delta t)-\vec{r}_j (t_w)))~,
\label{eq:ISF}
\end{equation}
}where the wave vector $\vec q$  has magnitude $q$.   The MSD is defined as:
 \begin{equation}
 R^2 (t_w,t_w+\Delta t) =\frac{1}{N} \sum_{j=1}^N |\vec{r}_j (t_w+\Delta t)-\vec{r}_j (t_w) |^2~.
\label{eq:MSD}
\end{equation} 
In a supercooled liquid, both the ISF and MSD should respect time-translational invariance, and should be independent of the waiting time, $t_w$.   From Figs. \ref{passive_MSD} and \ref{passive_ISF},  this expectation is met for densities $\rho < 0.1625$.   In this supercooled regime, we can fit the long-time decays of  the ISF (Eq. \ref{eq:ISF}), averaged over a range of $t_w$ from $\textrm{200,000 to 800,000}$,   to a stretched exponential form \cite{kob1995testing,berthier2007monte},  $ F_{q}(\Delta t)  \propto \exp(-(\Delta t/\tau_{\alpha})^{\beta})$, as shown in  Fig. \ref{ISF_stretched_fits} (a).  The $\tau_{\alpha}$ extracted from this fit increases by two orders of magnitude over the density range $0.12$ to $0.16$ (Fig. \ref{ISF_stretched_fits} (b)).
%We extract the $\alpha$ relaxation time, measure of the  the longest structural relaxation time in the liquid, by fitting the long-time decay of the ISF to a stretched-exponential form, as shown in Fig. \ref{ISF_stretched_fits}.   We find that the relaxation time increases by two orders of magnitude over the density range $0.12$ to $0.16$.
%%%%%%%%%%%%%%%%%%%%%%%%%%%%%%%%%%%%%%%%%%%%%%%%%%%%%%%
%%%%%%%%%%%%%%%%%%%%%%%%%%%%%%%%%%%%%%%%%%%%%%%%%%%%%%%

%%%%%%%%%%%%%%%%%%%%%%%%%%%%%%%%%%%%%%%%%%%%%%%%%%%%%%%
%%%%%%%%%%%%%%%%%%%%%%%%%%%%%%%%%%%%%%%%%%%%%%%%%%%%%%%
%%%%%%%%%%%%%%%%%%%%%%%%%%%%%%%%%%%%%%%%%%%%%%%%%%%%%%%
%%%%%%%%%%%%%%%%%%%%%%%%%%%%%%%%%%%%%%%%%%%%%%%%%%%%%%%

%%%%%%%%%%%%%%%%%%%%%%%%%%%%%%%%%%%%%%%%%%%%%%%%%%%%%%%
%%%%%%%%%%%%%%%%%%%%%%%%%%%%%%%%%%%%%%%%%%%%%%%%%%%%%%%
%%%%%%%%%%%%%%%%%%%%%%%%%%%%%%%%%%%%%%%%%%%%%%%%%%%%%%%
%%%%%%%%%%%%%%%%%%%%%%%%%%%%%%%%%%%%%%%%%%%%%%%%%%%%%%%

%%%%%%%%%%%%%%%%%%%%%%%%%%%%%%%%%%%%%%%%%%%%%%%%%%%%%%%
%%%%%%%%%%%%%%%%%%%%%%%%%%%%%%%%%%%%%%%%%%%%%%%%%%%%%%%
%%%%%%%%%%%%%%%%%%%%%%%%%%%%%%%%%%%%%%%%%%%%%%%%%%%%%%%
%%%%%%%%%%%%%%%%%%%%%%%%%%%%%%%%%%%%%%%%%%%%%%%%%%%%%%%

%%%%%%%%%%%%%%%%%%%%%%%%%%%%%%%%%%%%%%%%%%%%%%%%%%%%%%%
%%%%%%%%%%%%%%%%%%%%%%%%%%%%%%%%%%%%%%%%%%%%%%%%%%%%%%%
\begin{figure}[t!]
\centering
\includegraphics[width= 0.9\linewidth]{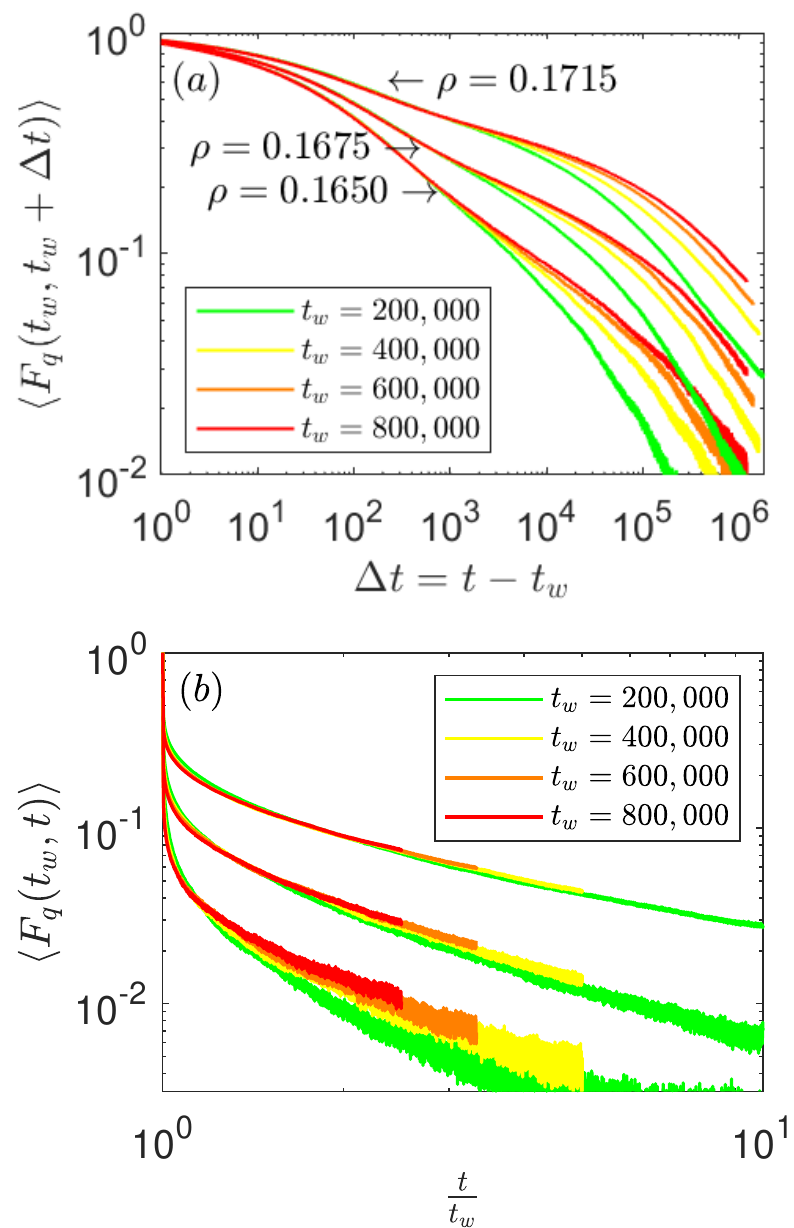}
\caption{Waiting time ($t_w$) dependence of the long time decay of the ISF at  $ \rho = 0.1650, 0.1675, 0.1715 $ (bottom to top).  The decay becomes slower with increasing $t_w$, and as shown in the inset,  depends only on the ratio $\frac{t}{t_w}$.}
\label{aging_figure} 
\end{figure}
%%%%%%%%%%%%%%%%%%%%%%%%%%%%%%%%%%%%%%%%%%%%%%%%%%%%%%%
%%%%%%%%%%%%%%%%%%%%%%%%%%%%%%%%%%%%%%%%%%%%%%%%%%%%%%%

Within our simulation time window of $ t_{\textrm{max}}= 2\times10^6$, passive hard crosses do not reach a time-translationally invariant state for densities $\ge 0.165$: both the ISF and the MSD depend on $t_w$,  at these densities.
We have established that in the aging regime the ISF exhibits a $t/t_w$ scaling, as shown in Fig. \ref{aging_figure}.  Scalings  of this form are characteristic of the aging regime in glassy systems~\cite{cugliandolo2002dynamics,kob1997aging,barrat1996ageing}.
% {\color{red} Need references to scaling in the aging regime and other glass references.  I am assuming that Carl has collected them for his thesis, and we can import them from his thesis bibliography.}

%%%%%%%%%%%%%%%%%%%%%%%%%%%%%%%%%%%%%%%%%%%%%%%%%%%%%%%
%%%%%%%%%%%%%%%%%%%%%%%%%%%%%%%%%%%%%%%%%%%%%%%%%%%%%%%

\vspace{0.5cm}
\section{Ensemble Averaged Density Distributions }
\label{appendix_b}
The standard evidence for MIPS is based on measurements of ensemble-averaged density distributions.  In this appendix, we present numerical results for these distributions over the full range of densities and activities.  These ensemble averages include both arrested and non-arrested states.   

%%%%%%%%%%%%%%%%%%%%%%%%%%%%%%%%%%%%%%%%%%%%%%%%%%%%%%%
%%%%%%%%%%%%%%%%%%%%%%%%%%%%%%%%%%%%%%%%%%%%%%%%%%%%%%%
\begin{figure}[t!]
\centering
\includegraphics[width= \linewidth]{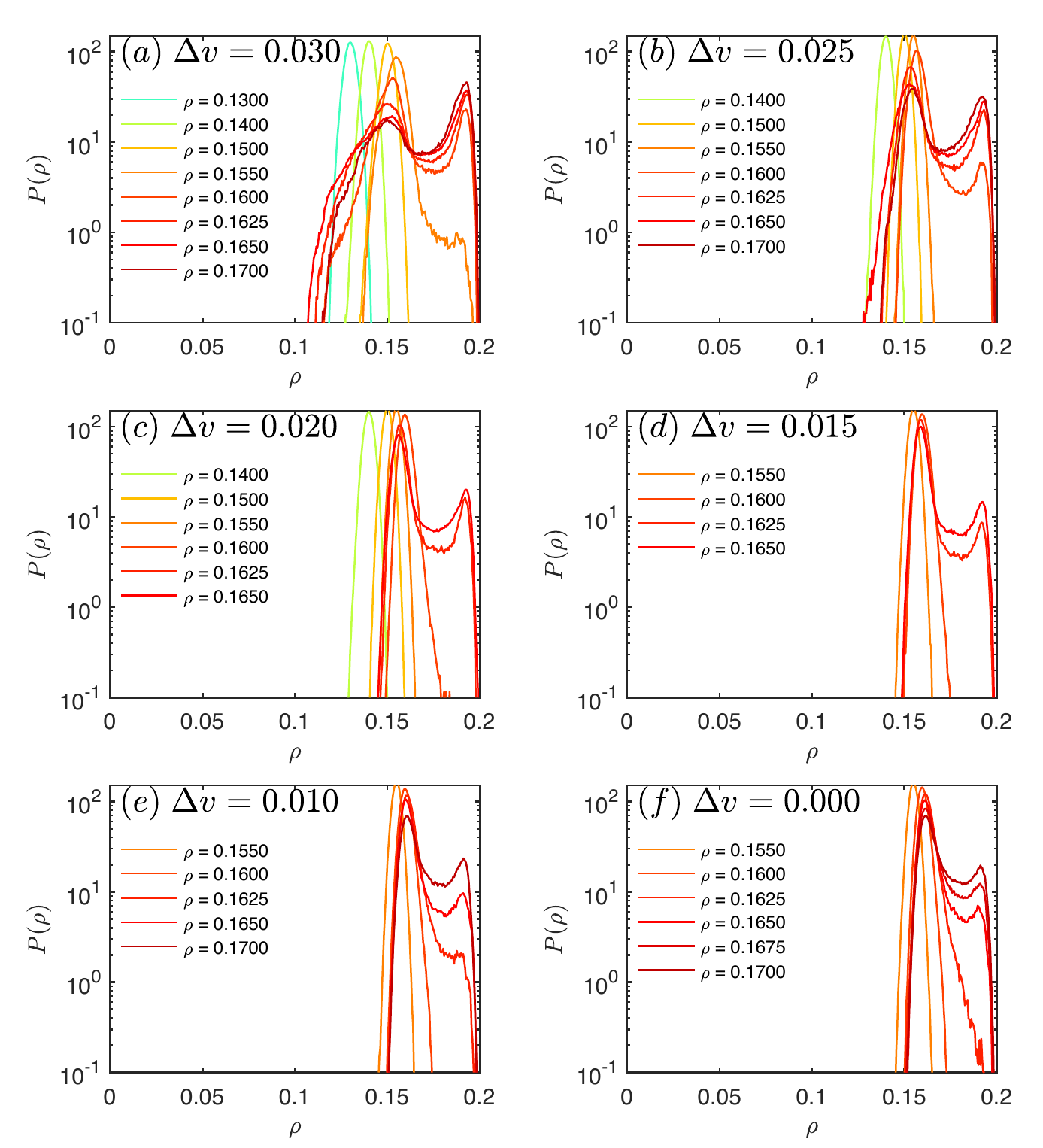}
\caption{Ensemble averaged local density distributions for smaller activity values showing the transition from the uniform liquid state to coexistence of solid-regions with liquid regions. The liquid peaks shows large low-density fluctuations for the largest activity values. For the passive system $ \Delta v = 0.000 $, a coexistence between fluid and solid regions appears for densities $ \rho > 0.1625 $. At non-zero but small values of the activity $ \Delta v > 0 $, this coexistence region expands to reach lower densities.}
\label{fig_density_distributions}
\end{figure}
%%%%%%%%%%%%%%%%%%%%%%%%%%%%%%%%%%%%%%%%%%%%%%%%%%%%%%%
%%%%%%%%%%%%%%%%%%%%%%%%%%%%%%%%%%%%%%%%%%%%%%%%%%%%%%%

As in standard equilibrium phase separation, MIPS phase boundaries are drawn based on coexisting peaks in the density distributions, whose positions do not depend on the global density but which trade intensity as the global density is varied~\cite{klamser2018thermodynamic}.  In our system, this expectation is met at low activities  (Fig. \ref{fig_density_distributions}) where we observe a continuation of the two-phase coexistence characteristic of the passive hard-crosses.

At higher activities, as seen in Fig. \ref{fig_density_distributions1},  the density distribution shows more structure. The ensemble averaged density distributions for the passive hard crosses exhibit a fluid-solid coexistence between $\rho \approx 0.16$ and $0.19$.   In the presence of  activity, the fluid peak at $\rho \approx 0.16$ becomes broader and ultimately develops a peak at very low densities, characteristic of the voids seen in the arrested states.  Before the void peak emerges clearly, in the solid-active liquid regime of the phase diagram,  the positions of these low-density peaks  are observed to depend both on activity and global density (c. f. Fig. \ref{fig_density_distributions1} (c)-(d)).  This observation is consistent with the fact that the phase separation is ``arrested''.   As we have shown using the ASEP-based coarse-grained theory, the arrest of this phase separation leads to a wetting layer of  width $\xi$ within which the active liquid is confined.  The density of particles in this active layer is a function of both activity and global density, as observed from the snapshots presented in the bottom row of Fig. \ref{2d_interface_samples}.

\vspace{0.5cm}
%%%%%%%%%%%%%%%%%%%%%%%%%%%%%%%%%%%%%%%%%%%%%%%%%%%%%%%
\section{Finite-Size Dependence of Phase Boundaries}
\label{appendix_d}
The coarse-grained theory, based on ASEP that we have used to  construct the non-equilibrium phase diagram~\cite{Cates_2004} of arrested states at $D_R=0$,  predicts that the phase boundaries depend explicitly on the system size $ L $.   In the infinite-system size limit, voids can form at any density and activity, according to this theory.  We have performed limited  studies of the size dependence of the phenomena we observe with the sole intent of checking whether the results of numerical simulations agree qualitatively with the predictions of the theory. 
 
In Fig~\ref{density_dist_finite_size} we compare ensemble averaged density distributions for system sizes $ L = 225, 450, \textrm{and} ~ L = 600$ at four different values of $\Delta v$ at $\rho=0.1500$. The results show that the activity at which phase separation commences decreases with increasing $L$.
%decreases that states which remain as simple, uniform active liquids in smaller systems are able to phase-separate, forming the acitve-liquid/solid phase, simply by making the entire system large enough. 
For example, at $\Delta v = 0.030$, the system is a homogeneous  fluid at  $ L = 225, 450 $, but there is fluid-solid coexistence for $ L = 600 $. Similarly, at $ \Delta v = 0.05$, the active-liquid solid phase is not  observed at $ L = 225 $. 

\begin{figure}[t!]
\centering
\includegraphics[width= 0.9\linewidth]{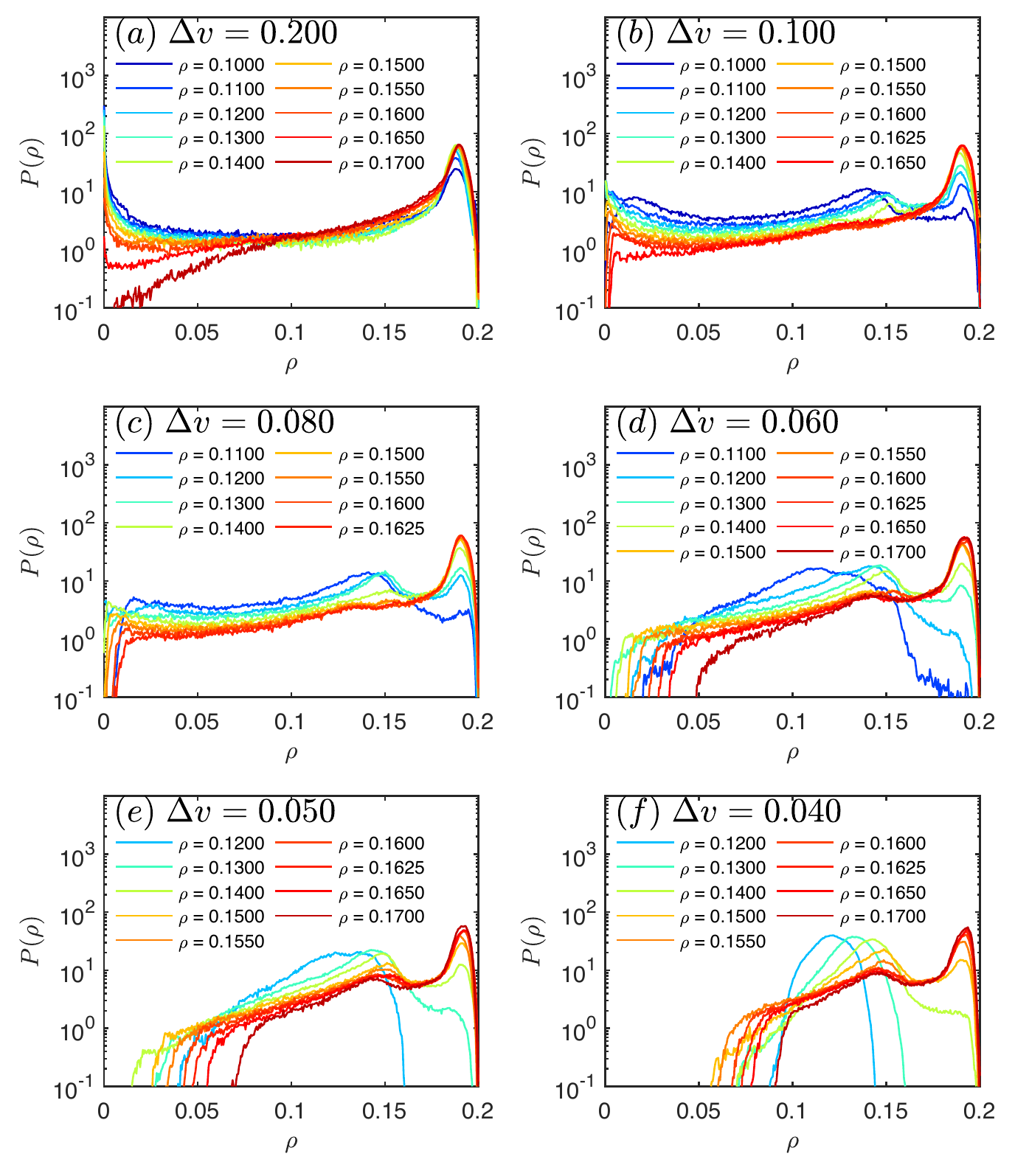}
\caption{Ensemble average distributions of the local density (measured in square boxes with side length $45$ lattice sites, for configurations at the simulation time $ t_{\textrm{max}}=2\times10^6$) compared for different activities $ \Delta v $ and different overall global densities $ \rho $. For large enough activity $ \Delta v \ge 0.2 $, there are only two clear peaks, a peak for the solid phase with density $ \rho_{\textrm{solid}} \approx 0.19 $ and a peak for the empty void regions $ \rho_{\textrm{void}} = 0 $. As the activity is reduced, a liquid-like peak near $ \rho \approx 0.14-0.15 $ appears. At small enough activity, the clear void peak disappears, and only large low density fluctuations of the liquid peak remain, suggesting that void regions have been filled in by the growing activity liquid interface. Note these distributions are ensemble averages over both arrested and non-arrested states.}
\label{fig_density_distributions1}
\end{figure}

\begin{figure}[t!]
\centering
\includegraphics[width= 0.9\linewidth]{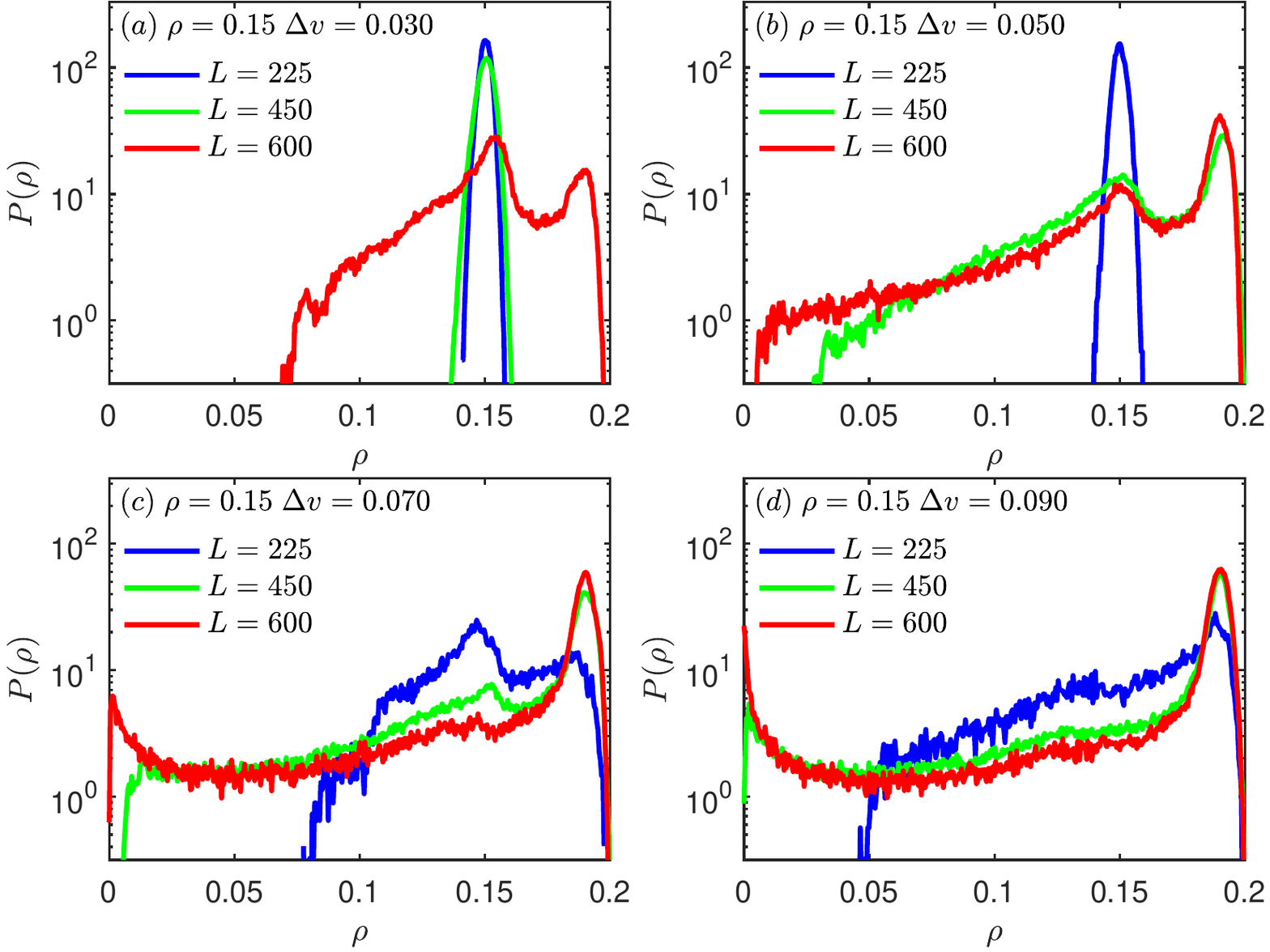}
\caption{Finite-size dependence of the local density distributions for several different activities at $ \rho = 0.15$ and linear system sizes $ L = 225, 450, 600 $.  The active liquid state (unimodal peak) present at $ \Delta v = 0.03 $ becomes a bimodal active liquid + solid coexistence for $L =600$. Similarly, at larger activity, for instance $ \Delta v = 0.07 $, increasing the system size opens up enough space for empty void regions to form, so that the active liquid + solid coexistence at $ L = 225 $ transforms into an active liquid + solid + void state at $ L = 600 $. The results qualitatively support the predicted finite size dependence of the boundaries between the three non-equilibrium phase types, active liquid, solid-active liquid, and solid-active liquid-void.}
\label{density_dist_finite_size}
\end{figure}

As mentioned above, the reason for the strong finite size effects is the ability of the system at $D_R=0$ to form voids at any activity and density.  This feature is validated in the simulations, as seen in Fig~\ref{density_dist_finite_size}, which shows regions that appear as low-density fluctuations in small systems transition to voids at larger $L$.   These effects are in turn a consequence of the emergence of a single length scale, $\xi$,  which depends only on $\Delta v$, characterizing the density variation.  When the system size is larger than $\xi$, voids open up to ensure mass conservation. Conversely, squeezing the system down to sizes smaller than $\xi$ forces the interfaces to overlap.   The ensemble averaged density distributions reflect this overlap through the development of a broad tail on the low density side of the solid peak  in the  the regime of solid-active liquid coexistence in {\it finite} systems.   Since this phase disappears in the thermodynamic limit, it cannot be characterized by thermodynamic measures such as binodals derived from the peaks of the density distributions. As seen in Fig. \ref{fig_density_distributions1}, the shape of the distribution that interpolates between the void and the solid peaks depends on the global density and the system size (Fig. \ref{density_dist_finite_size}).
%%%%%%%%%%%%%%%%%%%%%%%%%%%%%%%%%%%%%%%%%%%%%%%%%%%%%%%

%%%%%%%%%%%%%%%%%%%%%%%%%%%%%%%%%%%%%%%%%%%%%%%%%%%%%%%
%%%%%%%%%%%%%%%%%%%%%%%%%%%%%%%%%%%%%%%%%%%%%%%%%%%%%%%
%%%%%%%%%%%%%%%%%%%%%%%%%%%%%%%%%%%%%%%%%%%%%%%%%%%%%%%
%%%%%%%%%%%%%%%%%%%%%%%%%%%%%%%%%%%%%%%%%%%%%%%%%%%%%%%

%%%%%%%%%%%%%%%%%%%%%%%%%%%%%%%%%%%%%%%%%%%%%%%%%%%%%%%
%%%%%%%%%%%%%%%%%%%%%%%%%%%%%%%%%%%%%%%%%%%%%%%%%%%%%%%
%%%%%%%%%%%%%%%%%%%%%%%%%%%%%%%%%%%%%%%%%%%%%%%%%%%%%%%
%%%%%%%%%%%%%%%%%%%%%%%%%%%%%%%%%%%%%%%%%%%%%%%%%%%%%%%

%%%%%%%%%%%%%%%%%%%%%%%%%%%%%%%%%%%%%%%%%%%%%%%%%%%%%%%
%%%%%%%%%%%%%%%%%%%%%%%%%%%%%%%%%%%%%%%%%%%%%%%%%%%%%%%
\begin{figure}[t!]
\centering
\includegraphics[width= 0.75\linewidth]{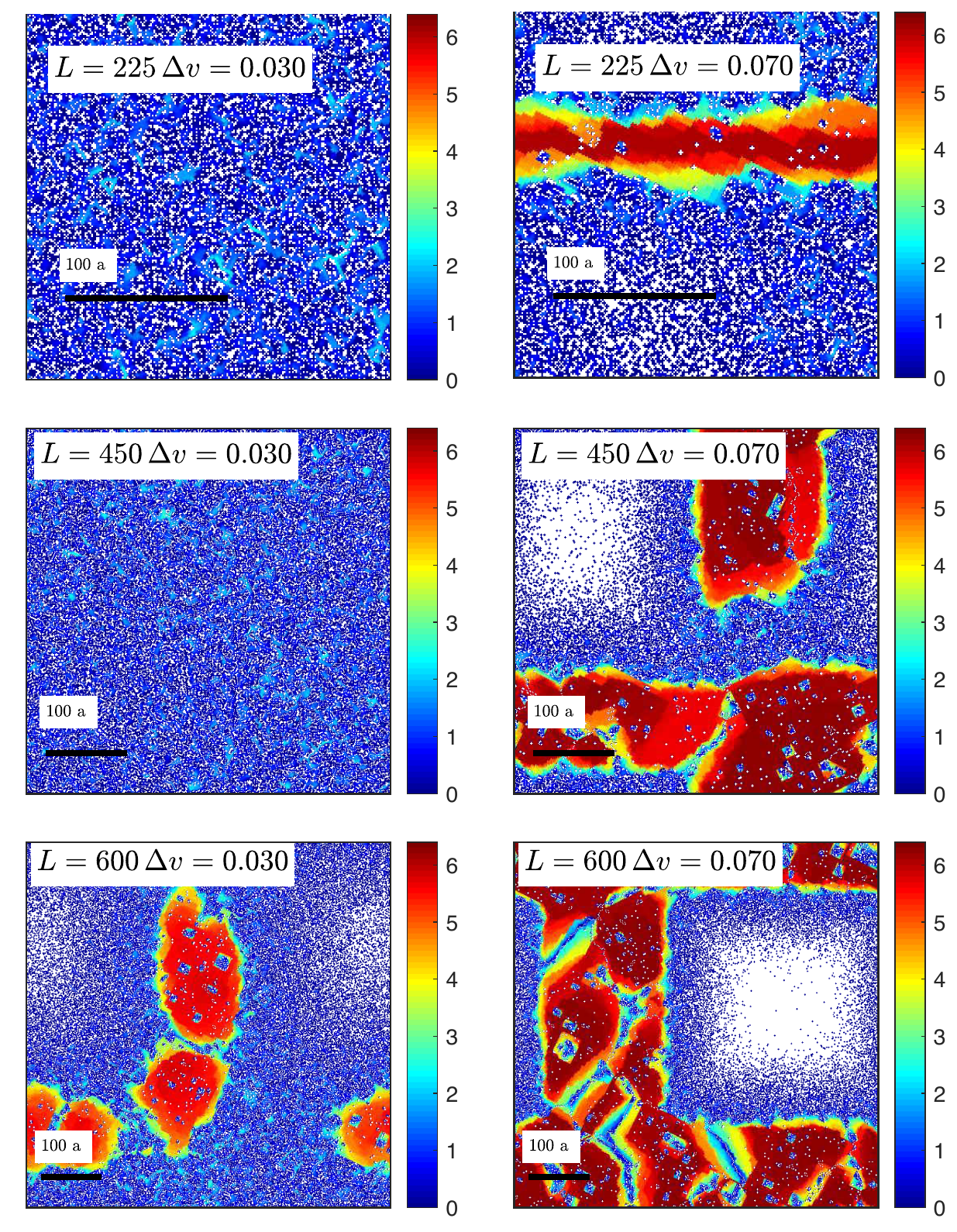}
\caption{Snapshots illustrating the finite-size dependence of the final non-equilibrium state reached, for $ \rho = 0.15$ and $ L = 225, 450, 600 $. For small activity $ \Delta v = 0.03$, increasing the system size transforms the final state from an active liquid in the smaller systems with $L =225, 450 $, to an active liquid + solid state in the largest system $ L = 600 $.  For larger activity $ \Delta v = 0.07 $, increasing the system size allows room for voids to open up, so that the active liquid + solid state at $ L = 225 $ clearly becomes an active liquid + solid + void state at $L= 600$. The black scale bars indicate a length of $ 100 $ lattice sites.}
\label{finite_size_pics}
\end{figure}
%%%%%%%%%%%%%%%%%%%%%%%%%%%%%%%%%%%%%%%%%%%%%%%%%%%%%%%
%%%%%%%%%%%%%%%%%%%%%%%%%%%%%%%%%%%%%%%%%%%%%%%%%%%%%%%

%%%%%%%%%%%%%%%%%%%%%%%%%%%%%%%%%%%%%%%%%%%%%%%%%%%%%%%
\begin{figure*}[h!]
  \includegraphics[width=1\textwidth]{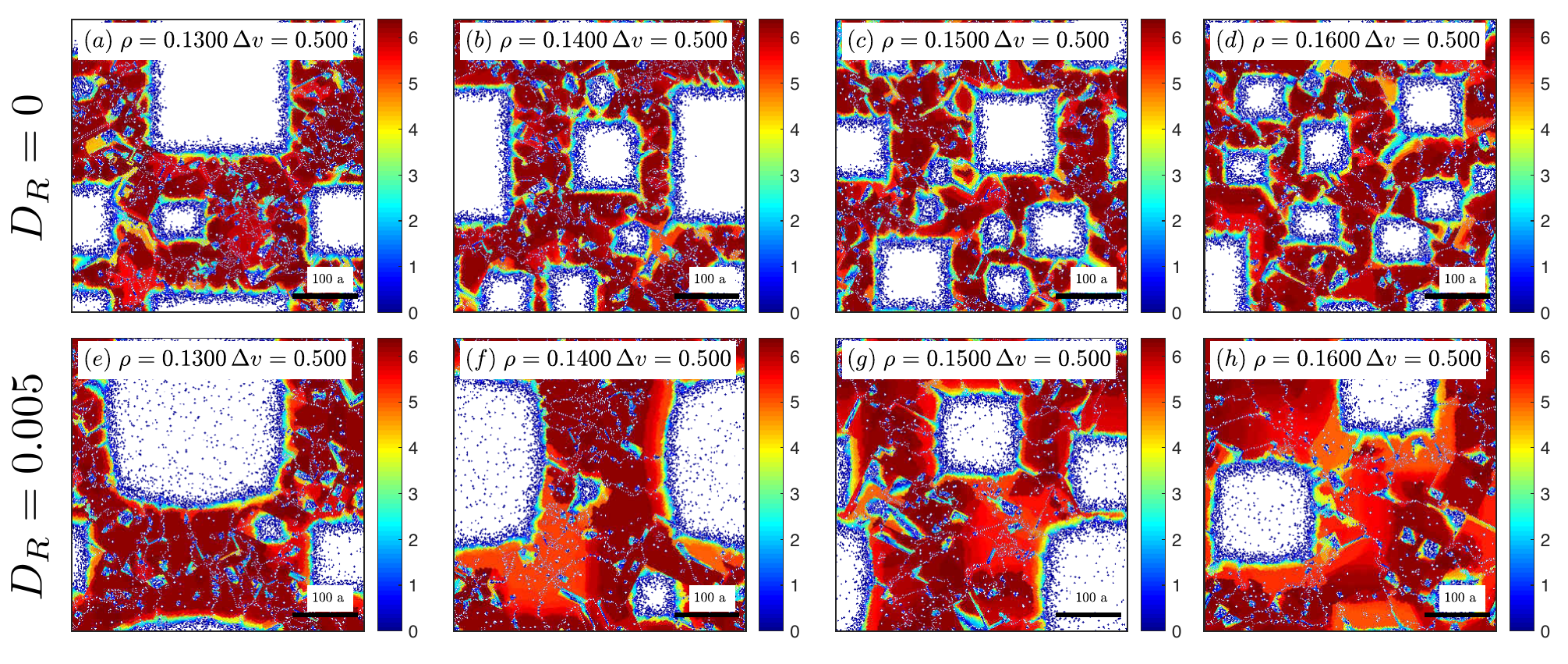} 
\caption{\label{finite_and_zero_rotation_comparison}  Comparison between the final gel-like configurations observed for the infinitely persistent $ D_R = 0 $ active crosses (a)-(d) and for finite rotation rate $ D_R = 0.005 $ (e)-(h). Color bars denote the stationary times of each cross, $ \log_{10}(\tau_i) $. For this large activity value $ \Delta v = 0.5 $, the percolating gel-like solid network can still form for active crosses with a finite rotation rate, suggesting that the rotational-locking mechanism of crosses greatly facilitates the global arrest of the system. For finite rotation rates, the rectangular empty void regions found for non-rotating crosses become filled with a finite density gas, and the boundaries of the low-density region become more rounded. The term ``gas" here refers to very dilute regions and ``liquid'' is used to the describe higher densities (no significant physical difference is meant to be implied).
}
\end{figure*}
%%%%%%%%%%%%%%%%%%%%%%%%%%%%%%%%%%%%%%%%%%%%%%%%%%%%%%%
%%%%%%%%%%%%%%%%%%%%%%%%%%%%%%%%%%%%%%%%%%%%%%%%%%%%%%%
\begin{figure*}[h!]
  \includegraphics[width=1\textwidth]{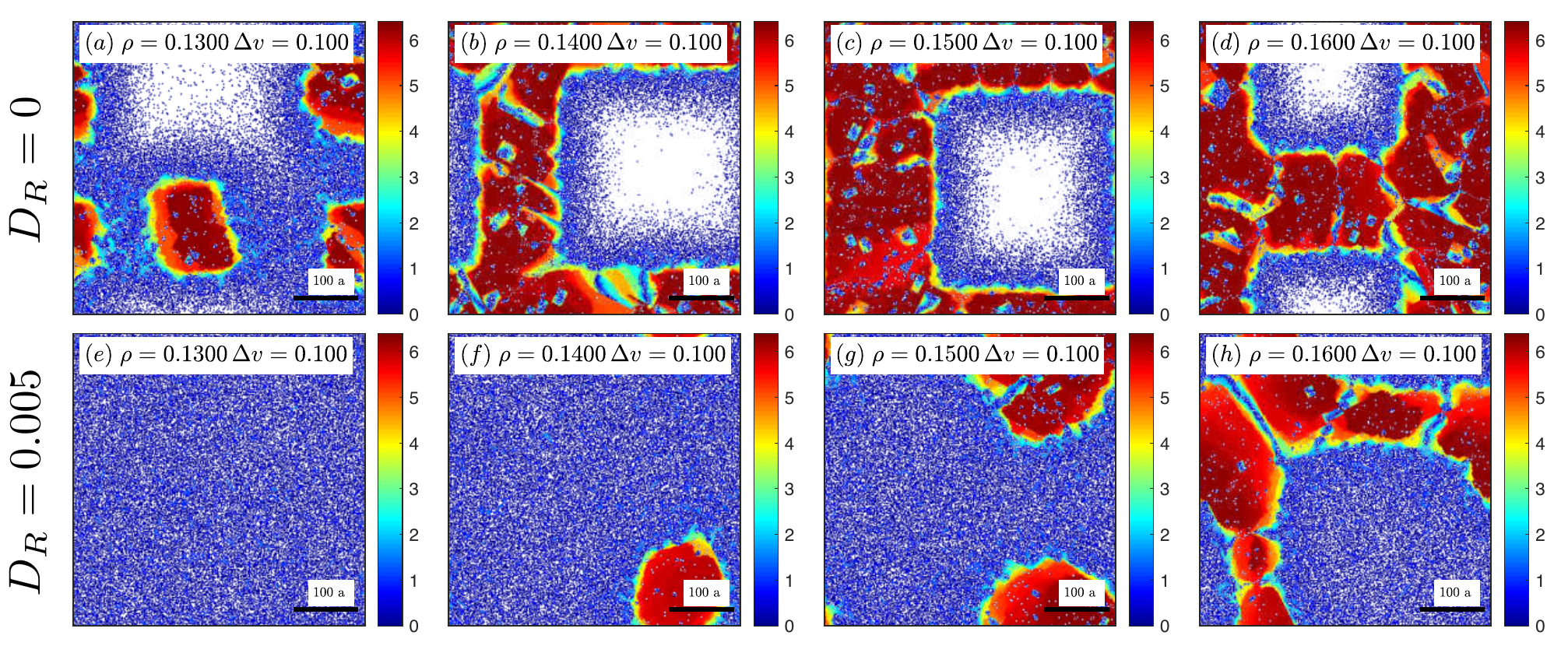} 
\caption{\label{finite_and_zero_rotation_comparison2}  Comparison of long time configurations for non-rotating and finite rotation rate $ D_R = 0.005 $ active crosses at a smaller activity value $ \Delta v = 0.100 $. Here, the non-rotating active crosses (a)-(d) separate into solid, void and intervening active liquid regions. The probability of forming a percolating solid network and becoming arrested increase with  increasing global density. For the finite rotation rate (e)-(h) a more standard two phase coexistence between liquid and solid is restored, and the structures  resemble a two-phase coexistence between a fluid and a solid with little density heterogeneity in the fluid phase.}
\end{figure*}
%%%%%%%%%%%%%%%%%%%%%%%%%%%%%%%%%%%%%%%%%%%%%%%%%%%%%%%

%
%Stronger qualitative differences emerge at weaker activities.  As seen in Figs. \ref{finite_and_zero_rotation_comparison2} and  \ref{density_dist_finite_and_zero_rot},  there is a more standard two-phase coexistence between a fluid and a solid, as observed in MIPS of hard spheres.  

%In agreement with 
%
%For example, in Fig.~\ref{density_dist_finite_and_zero_rot}(c), the distributions for the zero-rotation system at activity $\Delta v = 0.1$ show that the system is in the void/active-liquid/solid phase, with clear density peaks appearing at zero density. In contrast, the density distributions taken at the same activity and densities in the finite-rotation limit appear in Fig.~\ref{density_dist_finite_and_zero_rot}(d), where the density distributions now seem to show just a fluid-solid phase coexistence. Notably, turning on a finite-rotation rate has entirely removed the presence of any strong-low density fluctuations of the liquid phase, which are always present in some form in the zero-rotation system. The snapshots of the final states for the finite-rotation system shown in Fig.~\ref{finite_and_zero_rotation_comparison2} look much more similar to the kind of phase separation seen in systems exhibiting standard MIPS. The snapshots shown single, compact dense solid clusters surrounded by uniform fluid phase. 

%% probably  too much to go into the phase diagram for finite rotation rates in the paper SI, which I have included in my dissertation. CARL

%%%%%%%%%%%%%%%%%%%%%%%%%%%%%%%%%%%%%%%%%%%%%%%%%%%%%%%

\clearpage

\end{appendix}

%\clearpage
%
%\begin{widetext}

%\bibliographystyle{revtex4-2} 
%\bibliography{Active_Hard_Crosses_BibFile}

\bibliography{Active_Hard_Crosses_BibFile}

\end{document}